\pdfoutput=1

\documentclass[sigplan,screen]{acmart}

\acmJournal{PACMPL}
\acmVolume{1}
\acmNumber{CONF} 
\acmArticle{1}
\acmYear{2019}
\acmMonth{1}
\acmDOI{} 
\startPage{1}

\usepackage{booktabs}   
\usepackage{subcaption} 

\usepackage{xspace}     
\usepackage{mathpartir} 
\usepackage{paralist}   
\usepackage{fancyvrb}   

\usepackage{flushend}

\newcommand{\mdquote}[1]{\text{"}{#1}\text{"}}

\newcommand{\syntaxLine}[4]{${#1}$ & ${#2}$ & ${#3}$ & $\textit{#4}$ }
\newcommand{\namedRuleform}[2]{{\small $\ruleform{#1}$} \quad {#2}}
\newcommand{\vdashNamedD} [1]{\vdash_{{\scriptscriptstyle \hspace{-1mm}\mathtt{#1}}}}

\definecolor{superlightgray}{gray}{0.90}
\definecolor{Blue}{rgb}{0.0, 0.14, 0.6}
\definecolor{Sepia}{rgb}{0.30, 0.00, 0.00} 
\definecolor{OliveGreen}{rgb}{0.22, 0.26, 0.12}
\definecolor{Violet}{rgb}{0.56, 0.0, 1.0}

\newcommand{\highlight}[1]{\setlength{\fboxsep}{2pt}\colorbox{superlightgray}{\ensuremath{#1}}}

\newcommand{\ruleform}[1]{\fbox{$#1$}}

\newcommand{\fv}   [1]{\mathit{fv}(#1)}

\newcommand{\keyof}      {\begin{color}{Blue}\normalfont{\textbf{of}}\end{color}}
\newcommand{\keycase}    {\begin{color}{Blue}\normalfont{\textbf{case}}\end{color}}
\newcommand{\keydata}    {\begin{color}{Blue}\normalfont{\textbf{data}}\end{color}}
\newcommand{\keytype}    {\begin{color}{Blue}\normalfont{\textbf{type}}\end{color}}
\newcommand{\keyfamily}  {\begin{color}{Blue}\normalfont{\textbf{family}}\end{color}}
\newcommand{\keylet}     {\begin{color}{Blue}\normalfont{\textbf{let}}\end{color}}
\newcommand{\keyin}      {\begin{color}{Blue}\normalfont{\textbf{in}}\end{color}}
\newcommand{\keyclass}   {\begin{color}{Blue}\normalfont{\textbf{class}}\end{color}}
\newcommand{\keyinstance}{\begin{color}{Blue}\normalfont{\textbf{instance}}\end{color}}
\newcommand{\keywhere}   {\begin{color}{Blue}\normalfont{\textbf{where}}\end{color}}

\newcommand{\keyaxiom}   {\begin{color}{Blue}\normalfont{\textbf{axiom}}\end{color}}

\newcommand{\blueforall}{{\begin{color}{Blue}\forall\end{color}}}
\newcommand{\metaforall}[1]{\text{for each}~#1.~}

\newcommand{\sepia}[1]{{\begin{color}{Sepia}{#1}\end{color}}}

\newcommand{\Int}{{\begin{color}{OliveGreen}\mathit{Int}\end{color}}}

\newcommand{\Nat}{{\begin{color}{OliveGreen}\mathit{Nat}\end{color}}}
\newcommand{\Zero}{{\begin{color}{OliveGreen}\mathit{Z}\end{color}}} 
\newcommand{\Succ}{{\begin{color}{OliveGreen}\mathit{S}\end{color}}} 

\newcommand{\Vector}{{\begin{color}{OliveGreen}\mathit{Vec}\end{color}}}
\newcommand{\VN}{{\begin{color}{OliveGreen}\mathit{VN}\end{color}}}
\newcommand{\VC}{{\begin{color}{OliveGreen}\mathit{VC}\end{color}}}

\newcommand{\Maybe}{{\begin{color}{OliveGreen}\mathit{Maybe}\end{color}}}

\newcommand{\Bool}{{\begin{color}{OliveGreen}\mathit{Bool}\end{color}}}

\newcommand{\Term}{{\begin{color}{OliveGreen}\mathit{Term}\end{color}}}
\newcommand{\Con}{{\begin{color}{OliveGreen}\mathit{Con}\end{color}}}
\newcommand{\Tup}{{\begin{color}{OliveGreen}\mathit{Tup}\end{color}}}

\newcommand{\Add}{{\begin{color}{OliveGreen}\mathit{Add}\end{color}}}

\newcommand{\Eq}{{\begin{color}{OliveGreen}\mathit{Eq}\end{color}}}
\newcommand{\Ord}{{\begin{color}{OliveGreen}\mathit{Ord}\end{color}}}
\newcommand{\Show}{{\begin{color}{OliveGreen}\mathit{Show}\end{color}}}

\newcommand{\Monad}{{\begin{color}{OliveGreen}\mathit{Monad}\end{color}}}
\newcommand{\Applicative}{{\begin{color}{OliveGreen}\mathit{Applicative}\end{color}}}
\newcommand{\Functor}{{\begin{color}{OliveGreen}\mathit{Functor}\end{color}}}

\newcommand{\Monoid}{{\begin{color}{OliveGreen}\mathit{Monoid}\end{color}}}
\newcommand{\Alternative}{{\begin{color}{OliveGreen}\mathit{Alternative}\end{color}}}

\newcommand{\unwords}{{\begin{color}{Sepia}\mathit{unwords}\end{color}}}
\newcommand{\hsShow}{{\begin{color}{Sepia}\mathit{show}\end{color}}} 

\newcommand{\append}{{\begin{color}{Sepia}\mathit{append}\end{color}}}
\newcommand{\cmp}{{\begin{color}{Sepia}\mathit{cmp}\end{color}}}
\newcommand{\eq}{{\begin{color}{Sepia}\mathit{eq}\end{color}}}
\newcommand{\ctx}{{\begin{color}{Sepia}\mathit{ctx}\end{color}}}

\newcommand{\aA}{{\begin{color}{Sepia}\mathit{a}\end{color}}}
\newcommand{\bB}{{\begin{color}{Sepia}\mathit{b}\end{color}}}
\newcommand{\cC}{{\begin{color}{Sepia}\mathit{c}\end{color}}} 
\newcommand{\dD}{{\begin{color}{Sepia}\mathit{d}\end{color}}}

\newcommand{\fF}{{\begin{color}{Sepia}\mathit{f}\end{color}}}

\newcommand{\kK}{{\begin{color}{Sepia}\mathit{k}\end{color}}}
\newcommand{\mM}{{\begin{color}{Sepia}\mathit{m}\end{color}}}
\newcommand{\nN}{{\begin{color}{Sepia}\mathit{n}\end{color}}}

\newcommand{\xX}{{\begin{color}{Sepia}\mathit{x}\end{color}}}
\newcommand{\yY}{{\begin{color}{Sepia}\mathit{y}\end{color}}}

\newcommand{\XS}{{\begin{color}{Sepia}\mathit{xs}\end{color}}}
\newcommand{\YS}{{\begin{color}{Sepia}\mathit{ys}\end{color}}}

\newcommand{\Coq}{{\rm {\sc Coq}}\xspace}

\newcommand{\SystemFC}{System F$_\text{C}$\xspace}
\newcommand{\SystemF}{System F\xspace}
\newcommand{\HM}{{\rm {\sc HM}}\xspace}

\newcommand{\opEq} {\texttt{==}}
\newcommand{\opAnd}{\texttt{\&\&}}

\newcommand{\Unit}{\mathit{(\hspace{1pt})}} 

\newcommand{\as}{\overline{a}}
\newcommand{\bs}{\overline{b}}

\newcommand{\ds}{\overline{d}}

\newcommand{\ts}{\overline{t}}

\newcommand{\xs}{\overline{x}}

\newcommand{\TC}{\texttt{TC}}     
\newcommand{\tyCon}  {\mathit{T}} 
\newcommand{\dataCon}{\mathit{K}} 
\newcommand{\tyFam}  {\mathit{F}} 

\newcommand{\monotype}{\tau}
\newcommand{\qualtype}{\rho}
\newcommand{\polytype}{\sigma}

\newcommand{\monotypes}{\overline{\monotype}}

\newcommand{\program}{\mathit{pgm}}
\newcommand{\clsDecl}{\mathit{cls}}
\newcommand{\insDecl}{\mathit{ins}}
\newcommand{\valDecl}{\mathit{val}}
\newcommand{\decl}{\mathit{decl}}
\newcommand{\decls}{\mathit{decls}}

\newcommand{\tyEnv}{\Gamma}
\newcommand{\programTheory}{\mathit{P}}

\newcommand{\iComma}{,_{\mbox{\tiny I}}}

\newcommand{\lComma}{,_{\mbox{\tiny L}}}

\newcommand{\constraint}      {\mathit{Q}} 
\newcommand{\supann}          {^\text{sup}}
\newcommand{\supconstraint}   {\constraint\supann}
\newcommand{\constraints}     {\overline{\constraint}} 
\newcommand{\supconstraints}  {\overline{\supconstraint}}
\newcommand{\constraintSet}   {\mathit{C}} 
\newcommand{\constraintScheme}{\mathit{S}} 
\newcommand{\supdict}         {d\supann}
\newcommand{\supdicts}        {\overline{\supdict}}

\newcommand{\annConstraint}      {\mathcal{Q}} 
\newcommand{\annConstraintSet}   {\mathcal{C}} 

\newcommand{\localTheory}   {\annConstraintSet_L}

\newcommand{\axiomSet}{\mathcal{A}}
\newcommand{\invertedAxiomSet}{\axiomSet_B}
\newcommand{\instanceAxiomSet}{\axiomSet_I}
\newcommand{\superAxiomSet}   {\axiomSet_S}

\newcommand{\tySubst}{\theta}
\newcommand{\evSubst}{\eta}

\newcommand{\bisTcClsDeclNoShade}[5]{{#1} \vdashNamedD{CLS} {#2} : {#3} ; {#4} \rightsquigarrow {#5}}

\newcommand{\genProject}[3]{\mathit{proj}^{#1}_{#2}({#3})}

\newcommand{\axiom}[2]{g^{#1}_{#2}} 

\newcommand{\matchCtx}{\mathbb{E}}              
\newcommand{\applyExprCtx}[2]{{#1}[{#2}]}

\newcommand{\mponens}{\mathit{mponens}}
\newcommand{\mpOneStep}[3]{\mponens({#1}, {#2}, {#3})}
\newcommand{\mpClosure}[3]{\mathit{closure}({#1}, {#2}, {#3})}

\newcommand{\scName}{{\small\textsc{ScClosure}}}
\newcommand{\scClosure}[2]{\scName({#1}, {#2})}

\newcommand{\invScName}{{\small\textsc{InvScClosure}}}
\newcommand{\invScClosure}[2]{\invScName({#1}, {#2})}

\newcommand{\equalities}{\mathit{E}}

\newcommand{\tcElabTm}[6]{{#1} \vdashNamedD{TM} {#2} : {#3} \highlight{\rightsquigarrow {#4}} \mid {#5} ; {#6}}
\newcommand{\tcElabTmNoShade}[6]{{#1} \vdashNamedD{TM} {#2} : {#3} \rightsquigarrow {#4} \mid {#5} ; {#6}}

\newcommand{\tcElabTy}[1]{\mathit{elab}_{{\scriptscriptstyle\mathtt{TY}}}({#1})}
\newcommand{\tcElabCt}[1]{\mathit{elab}_{{\scriptscriptstyle\mathtt{CT}}}({#1})}

\newcommand{\unifyName}{\mathit{unify}}
\newcommand{\tcUnify}[2]{\unifyName({#1} ; {#2})}

\newcommand{\tcSimplifyCt}   [5]{{#1}; {#2} \models {#3} \rightsquigarrow {#4}\highlight{; {#5}}} 
\newcommand{\tcSimplifyAllCt}[5]{{#1}; {#2} \models {#3} \rightsquigarrow {#4}\highlight{; {#5}}} 

\newcommand{\tcSimplifyCtNoShade}[5]{{#1}; {#2} \models {#3} \rightsquigarrow {#4}; {#5}} 
\newcommand{\tcSimplifyAllCtNoShade}[5]{{#1}; {#2} \models {#3} \rightsquigarrow {#4}; {#5}} 

\newcommand{\tcElabIns}[5]{{#1}; {#2} \vdashNamedD{INS} {#3} : {#4} \highlight{\rightsquigarrow {#5}}}
\newcommand{\tcElabVal}[5]{{#1}; {#2} \vdashNamedD{VAL} {#3} : {#4} \highlight{\rightsquigarrow {#5}}}

\newcommand{\tcElabValNoShade}[5]{{#1}; {#2} \vdashNamedD{VAL} {#3} : {#4} \rightsquigarrow {#5}}
\newcommand{\tcElabInsNoShade}[5]{{#1}; {#2} \vdashNamedD{INS} {#3} : {#4} \rightsquigarrow {#5}}

\newcommand{\fcCast}[2]{{#1} \triangleright {#2}}

\newcommand{\tyFamApp}[2]{{#1}({#2})}

\newcommand{\fcType}{\upsilon}                
\newcommand{\fcTypes}{\overline{\fcType}}     

\newcommand{\fcEqCt}{\phi}                             
\newcommand{\fcEqCs}{\overline{\fcEqCt}}               

\newcommand{\coercion}{\gamma}
\newcommand{\coercions}{\overline{\coercion}}

\newcommand{\fsym}  [1]{\text{sym}~{#1}}
\newcommand{\frefl} [1]{\langle{#1}\rangle}
\newcommand{\fleft} [1]{\text{left}~{#1}}
\newcommand{\fright}[1]{\text{right}~{#1}}

\newcommand{\fcmp}{\mathbin{\raise 0.6ex\hbox{\oalign{\hfil$\scriptscriptstyle \mathrm{o}$\hfil\cr\hfil$\scriptscriptstyle\mathrm{9}$\hfil}}}}
\newcommand{\fcomp}[2]{{#1}\fcmp{#2}}
\newcommand{\finst}[2]{{#1}[{#2}]}
\newcommand{\finstQual}[2]{{#1}@{#2}}

\newcommand{\fcTcTy}[2]{{#1} \vdashNamedD{\bf ty} {#2}} 
\newcommand{\fcTcPr}[2]{{#1} \vdashNamedD{\bf pr} {#2}}
\newcommand{\fcTcCo}[3]{{#1} \vdashNamedD{\bf co} {#2} : {#3}}
\newcommand{\fcTcTm}[3]{{#1} \vdashNamedD{\bf tm} {#2} : {#3}}
\newcommand{\fcTcPat}[3]{{#1} \vdashNamedD{\bf pat} {#2} : {#3}}
\newcommand{\fcTcDecl}[3]{{#1} \vdashNamedD{\bf decl} {#2} : {#3}}

\newcommand{\covar}{\omega}

\newcommand{\fcPat}{\mathit{p}}               

\newcommand{\tyPat}{\mathit{u}}
\newcommand{\tyPats}{\overline{\tyPat}}

\newcommand{\fcEnv}{\Delta}

\newcommand{\tcTcCt}[3]{{#1} \vdashNamedD{CT} {#2} \highlight{\rightsquigarrow {#3}}}
\newcommand{\tcTcTy}[3]{{#1} \vdashNamedD{TY} {#2} \highlight{\rightsquigarrow {#3}}} 
\newcommand{\tcTcTm}[5]{{#1}; {#2} \vdashNamedD{TM} {#3} : {#4} \highlight{\rightsquigarrow {#5}}}
\newcommand{\tcEntailCt}[4]{{#1}; {#2} \models \highlight{{#3} :} {#4}}

\newcommand{\tcTcClsDecl}[5]{{#1} \vdashNamedD{CLS} {#2} : {#3}; {#4} \highlight{\rightsquigarrow {#5}}}
\newcommand{\tcTcInsDecl}[5]{{#1}; {#2} \vdashNamedD{INS} {#3} : {#4} \highlight{\rightsquigarrow {#5}}}
\newcommand{\tcTcValDecl}[5]{{#1}; {#2} \vdashNamedD{VAL} {#3} : {#4} \highlight{\rightsquigarrow {#5}}}

\newcommand{\tcSubsumption}[6]{{#1}; {#2}; {#3} \vdashNamedD{TM} {#4} : {#5} \highlight{\rightsquigarrow {#6}}}

\newcommand{\tcTcCtNoShade}[3]{{#1} \vdashNamedD{CT} {#2} \rightsquigarrow {#3}}
\newcommand{\tcTcTyNoShade}[3]{{#1} \vdashNamedD{TY} {#2} \rightsquigarrow {#3}} 
\newcommand{\tcTcTmNoShade}[5]{{#1}; {#2} \vdashNamedD{TM} {#3} : {#4} \rightsquigarrow {#5}}
\newcommand{\tcEntailCtNoShade}[4]{{#1}; {#2} \models {#3} : {#4}}
\newcommand{\tcTcClsDeclNoShade}[5]{{#1} \vdashNamedD{CLS} {#2} : {#3}; {#4} \rightsquigarrow {#5}}
\newcommand{\tcTcInsDeclNoShade}[5]{{#1}; {#2} \vdashNamedD{INS} {#3} : {#4} \rightsquigarrow {#5}}

\newcommand{\tcSubsumptionNoShade}[6]{{#1}; {#2}; {#3} \vdashNamedD{TM} {#4} : {#5} \rightsquigarrow {#6}}

\makeatletter
\newcommand\footnoteref[1]{\protected@xdef\@thefnmark{\ref{#1}}\@footnotemark}
\makeatother

\setcopyright{acmcopyright}
\copyrightyear{2019}
\acmYear{2019}
\acmConference[Haskell '19]{Proceedings of the 12th ACM SIGPLAN International Haskell Symposium}{August 22--23, 2019}{Berlin, Germany}
\acmBooktitle{Proceedings of the 12th ACM SIGPLAN International Haskell Symposium (Haskell '19), August 22--23, 2019, Berlin, Germany}
\acmPrice{15.00}
\acmDOI{10.1145/3331545.3342596}
\acmISBN{978-1-4503-6813-1/19/08}

\begin{document}

\title[Bidirectional Type Class Instances (Extended Version)] 
      {Bidirectional Type Class Instances (Extended Version)} 





\author{Koen Pauwels}
\affiliation{
  \institution{KU Leuven}
  \country{Belgium}
}
\email{koen.pauwels@cs.kuleuven.be}

\author{Georgios Karachalias}
\affiliation{
  \institution{KU Leuven}
  \country{Belgium}
}
\email{gdkaracha@gmail.com} 

\author{Michiel Derhaeg}
\affiliation{
  \institution{Guardsquare}
  \country{Belgium}
}
\email{michiel@derhaeg.be}

\author{Tom Schrijvers}
\affiliation{
  \institution{KU Leuven}
  \country{Belgium}
}
\email{tom.schrijvers@cs.kuleuven.be}

\renewcommand{\shortauthors}{K. Pauwels, G. Karachalias, M. Derhaeg, and T. Schrijvers}

\begin{abstract}
GADTs were introduced in Haskell's eco-system more than a decade ago, but their
interaction with several mainstream features such as type classes and
functional dependencies has a lot of room for improvement. More specifically,
for some GADTs it can be surprisingly difficult to provide an instance for even
the simplest of type classes.

In this paper we identify the source of this shortcoming and address it by
introducing a conservative extension to Haskell's type classes: {\em
Bidirectional Type Class Instances}. In essence, under our interpretation class
instances correspond to logical bi-implications, in contrast to their
traditional unidirectional interpretation.

We present a fully-fledged design of bidirectional instances, covering the
specification of typing and elaboration into System FC, as well as an algorithm for type
inference and elaboration. We provide a proof-of-concept
implementation of our algorithm, and revisit the meta-theory of type classes in
the presence of our extension.
%
\end{abstract}



\maketitle

\section{Introduction}
\label{sec:classes-bidirectional:intro}

Type classes were first introduced by~\citet{adhoc-polymorphism} as a
principled way to support ad-hoc polymorphism in Haskell, have since appeared
in other declarative languages like \Coq~\citep{Coq:manual} and
Mercury \citep{mercury}, and have influenced the design of similar
features (e.g., concepts for C++~\citep{concepts-article}).

One of the main reasons type classes have been so successful is that they
support sound, decidable, and efficient type inference~\citep{qualtypes},
while being a simple extension of the well-understood Hindley-Damas-Milner
system (HM)~\citep{hindley,DamasMilner}. Furthermore,
as~\citet{adhoc-polymorphism} have shown, they can be straightforwardly
translated to parametric polymorphism in intermediate languages akin to
\SystemF~\citep{girardthesis,reynolds-systemf-1,reynolds-systemf-2}.

Since the conception of type classes, instances have been interpreted as
logical implications, due to their straightforward implementation as \SystemF
functions. For example, the well-known equality instance for lists
\[
\keyinstance~\Eq~\aA \Rightarrow \Eq~[\aA]
\]
can be read as {\em`` if $\aA$ is an instance of $\Eq$, then so is $[\aA]$''}.
This interpretation has worked pretty well so far, but falls short in the
presence of advanced features such as {\em Generalized Algebraic Data Types}
(GADTs)~\citep{PeytonJones06}.

More specifically, with the current interpretation of type classes a large
class of GADTs cannot be made an instance of the simplest of type classes.
In this work we alleviate this problem by introducing a conservative
extension\footnote{By ``conservative'', we mean that our system can type
strictly more programs than plain Haskell does.}
to
Haskell: {\em Bidirectional Instances}. Under our interpretation, instances
like the above can be read as {\em ``$\aA$ is an instance of $\Eq$ {\bf if and
    only if} $[\aA]$ is an instance of $\Eq$''}.\footnote{
  Which we believe reflects what Haskell users already have in mind. Indeed,
  most prior research on type classes---such as the work of~\citet{fundeps-chr}---treat $\Eq~\aA$ and $\Eq~[\aA]$ as denotationally equivalent.
}

The main problem is that, unlike
basic type classes, this extension requires a non-parametric encoding (which
possibly explains its initial omission).
We overcome this problem with \SystemFC coercions~\citep{systemfc} and add a non-parametric
witness for the instance context to the otherwise parametric dictionary representation.
Our specific contributions are:

\begin{itemize}
  \item
    We present a detailed overview of the shortcomings in the interaction
    between GADTs and type classes, as well as other class-based extensions
    (Sec.~\ref{sec:classes-bidirectional-motivation}).
  \item
    We identify the major challenges of interpreting and elaborating type class
    instances bidirectionally (Sec.~\ref{sec:classes-bidirectional-challenges}).
  \item
    We describe an elaboration strategy that addresses all such challenges,
    while making Haskell strictly more expressive
    (Sec.~\ref{sec:classes-bidirectional-formal-extensions}).
  \item
    We formalize superclasses, an important---yet often neglected---aspect of
    type classes. Our formalization includes a specification of typing,
    elaboration, and an algorithm for type inference with elaboration
    (Sec.~\ref{sec:basic-with-superclasses}).
  \item
    We provide a formalization of typing and evidence translation from source
    terms to \SystemFC for type classes with bidirectional instances, as well
    as an algorithm for type inference with elaboration
    (Sec.~\ref{sec:the-basic-system}). Our approach reuses most of the
    infrastructure needed by superclasses; existing systems need to be
    minimally extended for the additional expressive power.
  \item
    We elaborate on the changes bidirectional instances induce to the
    meta-theory of type classes; notably termination of type inference and
    principality of types (Sec.~\ref{sec:classes-bidirectional-metatheory}).
  \item
    We provide a prototype implementation of our algorithm for type inference with
    elaboration at \url{https://github.com/KoenP/bidirectional-instances}.
\end{itemize}

\section{Motivation}\label{sec:classes-bidirectional-motivation}

\subsection{Structural Induction Over Indexed Data Types}
\label{sec:structural-induction}

Ever since GADTs were introduced in Haskell~\citep{PeytonJones06}, they have
been put to good use by programmers for dataflow analysis and
optimization~\citep{hoopl}, accelerated array
processing,\footnote{\url{https://hackage.haskell.org/package/accelerate}}
automatic
differentiation,\footnote{\url{https://hackage.haskell.org/package/ad}} and
much more.
Yet, their interaction with existing features such as type
classes~\citep{adhoc-polymorphism} and functional
dependencies~\citep{fundeps-original} can lead to surprising problems.
%

For example, consider (a simplified version of) the $\Term$ datatype, as
given by~\citet{FoundationsGADTs}:
\[
\begin{array}{l@{\hspace{1mm}}c@{\hspace{1mm}}l}
\multicolumn{3}{l}{\keydata~\Term :: \star \to \star~\keywhere} \\
  \hspace{2mm}\Con & :: & \aA \to \Term~\aA \\
  \hspace{2mm}\Tup & :: & \Term~\bB \to \Term~\cC \to \Term~(\bB,\cC) \\
\end{array}
\]
The GADT $\Term$ encodes a simple expression language, with constants
(constructed by data constructor $\Con$) and tuples (constructed by
data constructor $\Tup$).

Making ($\Term~\aA$) an instance of even the simplest of type classes can
be challenging. For example, the following straightforward instance is not
typeable under the current specification of type classes:
\[
\begin{array}{l@{\hspace{1mm}}l@{\hspace{1mm}}c@{\hspace{1mm}}l}
\multicolumn{4}{l}{\keyinstance~\Show~\aA \Rightarrow \Show~(\Term~\aA)~\keywhere} \\
  \hspace{2mm}\hsShow & (\Con~\xX)   & = & \hsShow~\xX \\
  \hspace{2mm}\hsShow & (\Tup~\xX~\yY) & = & \unwords~[\mdquote{(},\hsShow~\xX,\mdquote{,},\hsShow~\yY,\mdquote{)}] \\
\end{array}
\]
Loading the above program into \texttt{ghci} emits the following error message:
{\bfseries
\begin{Verbatim}[fontsize=\footnotesize]
 Bidirectional.hs:14:33:
     Could not deduce (Show b) arising from a use of `show'
     from the context (Show a) or from (a ~ (b, c))
\end{Verbatim}
\begin{Verbatim}[fontsize=\footnotesize]
 Bidirectional.hs:14:44:
     Could not deduce (Show c) arising from a use of `show'
     from the context (Show a) or from (a ~ (b, c))
\end{Verbatim}
}
\noindent
As the message indicates, the sources of the errors are the two recursive calls to
$\hsShow$ in the second clause: the instance context $(\Show~\aA)$
and the local constraint (exposed via GADT pattern matching) $\aA \sim (\bB, \cC)$
are not sufficient to prove ($\Show~\bB$) and ($\Show~\cC$), which are
required by the recursive calls to $\hsShow$. In summary, the type system
cannot derive the following implications:
\[
\begin{array}{l@{\hspace{1mm}}c@{\hspace{1mm}}l}
  \blueforall \bB.~\blueforall \cC.~\Show~(\bB, \cC) & \Rightarrow & \Show~\bB \\
  \blueforall \bB.~\blueforall \cC.~\Show~(\bB, \cC) & \Rightarrow & \Show~\cC \\
\end{array}
\]

\paragraph{The Problem}

Both implications above constitute the inversion of the implication derived by
the predefined $\Show$ instance for tuples:
\[
\keyinstance~(\Show~\bB, \Show~\cC) \Rightarrow \Show~(\bB,\cC)~\keywhere~\{~\ldots~\}
\]
Indeed, the interpretation of type classes in
existing systems is not {\em
bidirectional}: the system can only derive $\Show~(\bB,\cC)$ from $\Show~\bB$ and
$\Show~\cC$, but not the other way around.

\subsection{Functional Dependencies and Associated Types}

Unfortunately, the lack of bidirectionality of type class instances does not
only affect the expressive power of simple type classes, but also the
expressive power of features based on them, such as functional
dependencies~\citep{fundeps-original} and associated type
synonyms~\citep{AssociatedTypeSynonyms}.

For example, let us consider an example of type-level programming using
functional dependencies.\footnote{A similar example has been presented
by~\citet{FunWithFDs}, who implemented insertion sort at the level of types
using functional dependencies.} First, we define type-level natural numbers and
length-indexed vectors:
\[
\begin{array}{@{\hspace{0mm}}l@{\hspace{5mm}}l@{\hspace{0mm}}}
  \begin{array}{@{\hspace{0mm}}l@{\hspace{1mm}}c@{\hspace{1mm}}l@{\hspace{0mm}}}
    \multicolumn{3}{@{\hspace{0mm}}l@{\hspace{0mm}}}{\keydata~\Nat :: \star~\keywhere} \\
    \hspace{2mm}\Zero & :: & \Nat \\
    \hspace{2mm}\Succ & :: & \Nat \to \Nat \\
  \end{array}
&
  \begin{array}{@{\hspace{0mm}}l@{\hspace{1mm}}c@{\hspace{1mm}}l@{\hspace{0mm}}}
    \multicolumn{3}{@{\hspace{0mm}}l@{\hspace{0mm}}}{\keydata~\Vector :: \Nat \to \star \to \star~\keywhere} \\
    \hspace{2mm}\VN & :: & \Vector~\Zero~\aA \\
    \hspace{2mm}\VC & :: & \aA \to \Vector~\nN~\aA \to \Vector~(\Succ~\nN)~\aA \\
  \end{array}
\end{array}
\]
On the left, we define type-level natural numbers $\Nat$. Type $\Nat$ is
automatically promoted into a kind and data constructors $\Zero$ and $\Succ$
into type constructors of the same name, using the GHC extension
\texttt{DataKinds}~\citep{hspromoted}.

Length-indexed vectors $\Vector$ utilize $\Nat$ to index data constructors
$\VN$ and $\VC$ with the appropriate length: $\VN$ represents the empty vector
(and thus has length $\Zero$), and $\VC$ represents concatenation (and thus
constructs vectors of length $(\Succ~\nN)$, where $\nN$ is the length of the
tail).

Equipped with type-level natural numbers, we can encode type-level addition
(using the~\citet{peano-axioms} axioms) by means of a multi-parameter
type class and a functional dependency:
\[
\begin{array}{@{\hspace{0mm}}l@{\hspace{1mm}}l@{\hspace{1mm}}c@{\hspace{1mm}}l@{\hspace{1mm}}l@{\hspace{1mm}}l@{\hspace{1mm}}l@{\hspace{0mm}}}
  \multicolumn{7}{@{\hspace{0mm}}l@{\hspace{0mm}}}{\keyclass~\Add~(\nN :: \Nat)~(\mM :: \Nat)~(\kK :: \Nat) \mid \nN~\mM \to \kK} \\
  \keyinstance &                  &             & \Add & \Zero       & \mM & \mM         \\
  \keyinstance & \Add~\nN~\mM~\kK & \Rightarrow & \Add & (\Succ~\nN) & \mM & (\Succ~\kK) \\
\end{array}
\]
Parameters $\nN$ and $\mM$ represent the operands, and parameter $\kK$
represents the result, which is uniquely determined by the choice of $\nN$ and
$\mM$. The two Peano axioms for addition correspond to two instances for class
$\Add$, one for each form $\nN$ can take.

The above can be combined to define function $\append$, which
concatenates two length-indexed vectors:
\[
\begin{array}{l@{\hspace{1mm}}l@{\hspace{1mm}}l@{\hspace{1mm}}c@{\hspace{1mm}}l}
  \multicolumn{5}{l}{\append :: \Add~\nN~\mM~\kK \Rightarrow \Vector~\nN~\aA \to \Vector~\mM~\aA \to \Vector~\kK~\aA} \\
  \append & \VN           & \YS & = & \YS \\
  \append & (\VC~\xX~\XS) & \YS & = & \VC~\xX~(\append~\XS~\YS) \\
\end{array}
\]
The implementation of $\append$ is identical to the corresponding one for
simple lists but its signature is much richer: $\append$ takes two vectors of
lengths $\nN$ and $\mM$, and computes a vector of length $\kK$, where $\nN+\mM
= \kK$. Types like those above are extremely useful, for example in linear algebra
libraries (see for example Hackage package
\href{https://hackage.haskell.org/package/linear}{\texttt{linear}}), to ensure
that operations respect the expected dimensions.

Unfortunately, examples like the one above are known not to type-check, mainly due
to the lack of an evidence-based translation of functional dependencies. Yet,
even with the recent advances of~\citet{fundeps-core}, the above program is
ill-typed.

Once again, the key missing element is bidirectional instances. In the second
clause of $\append$, the recursive invocation of $\append$ requires
$(\Add~\nN'~\mM~\kK')$, while the signature provides
$(\Add~(\Succ~\nN')~\mM~(\Succ~\kK'))$.\footnote{In fact, the signature
provides $(\Add~\nN~\mM~\kK)$, which we can refine using $\nN \sim \Succ~\nN'$
(obtained by GADT pattern matching), and the type-level function introduced by
the functional dependency~\citep{fundeps-core}.} That is, we need the following
implication:
\[
\blueforall \nN'~\mM~\kK'.~\Add~(\Succ~\nN')~\mM~(\Succ~\kK') \Rightarrow \Add~\nN'~\mM~\kK'
\]
which can be obtained by interpreting the second $\Add$ instance
bidirectionally.

As has been speculated by many and has recently been illustrated
by~\citet{fundeps-core}, associated type
synonyms~\citep{AssociatedTypeSynonyms} share---for the most part---the same
semantics with functional dependencies. Thus, the problem we are presenting
here applies to associated type families as well; shortcomings of type classes
affect all their extensions (indeed, rewriting the above example to use
associated type synonyms instead of functional dependencies does not obviate
the need for bidirectionality).

In summary, the lack of bidirectionality of type class instances severely
reduces the expressive power of type class extensions, such as associated
types~\citep{DataFamilies}, associated type
synonyms~\citep{AssociatedTypeSynonyms}, and functional
dependencies~\citep{fundeps-original}.

\subsection{Summary}

In summary, the lack of a bidirectional elaboration of class instances
seriously undermines the interaction between GADTs and type classes, as well as
type class extensions. This is precisely the issue we address in this paper.

\section{Technical Challenges}\label{sec:classes-bidirectional-challenges}

Though bidirectional instances are sorely needed for applications involving
GADTs, the problem is more general. For example, no Haskell compiler accepts
programs where $\Eq~\aA$ needs to be derived from $\Eq~[\aA]$. This is the case
for the following type-annotated function:
\footnote{
  The language used for our examples is equivalent to Haskell 98 plus
  the \texttt{FlexibleContexts} and \texttt{GADTs} extensions.
}
\[
\begin{array}{l}
  \cmp :: \Eq~[\aA] \Rightarrow \aA \to \aA \to \Bool \\
  \cmp~\xX~\yY = \xX~\opEq~\yY \\
\end{array}
\]
Though contrived, function $\cmp$ is a minimal example that exhibits all
problems that arise in elaborating class instances bidirectionally in the
well-established dictionary-passing translation~\citep{ClassElaboration}. Thus,
we use it as our running example throughout the remainder of this section to
discuss the technical challenges of interpreting type class instances
bidirectionally.

\subsection{Key Idea}\label{sec:classes-bidirectional-challenges:key-idea}

\paragraph{Why Are Instances Bidirectional}


Existing systems with type classes ensure {\em coherence}\footnote{Elaboration
is said to be coherent if all valid typing derivations for a given program lead
to target programs with the same dynamic semantics.} by disallowing instance
heads to overlap. In turn, if no instance heads overlap, there can be at most
one derivation for making a concrete type an instance of a certain type class.
For example, given that instances are non-overlapping, the only way one can
derive $\Eq~[\Int]$ is by combining the following two instances:
\[
\begin{array}{l}
  \keyinstance~\Eq~\Int \\
  \keyinstance~\Eq~\aA \Rightarrow \Eq~[\aA] \\
\end{array}
\]
That being said, the only way one can create a dictionary of type $\Eq~[\aA]$,
for {\em any type} $\aA$, is by using the $\Eq$ instance for lists.
Consequently, if a constraint $\Eq~[\aA]$ is {\em given}, one can safely assume
that $\Eq~\aA$ is also available: modus ponens is invertible if there is no
overlap in the implication heads.

\paragraph{General Strategy}

In order to integrate bidirectionality in the system, we need to show how to
derive the instance context from the instance head {\em constructively}.
To achieve this, our strategy is
simple: {\em reuse the approach of superclasses}.

According to the traditional dictionary-passing translation of type
classes~\citep{ClassElaboration}, superclass dictionaries are stored within
subclass dictionaries. Hence, a superclass constraint (e.g., $\Eq~\aA$) can
always be derived from a subclass constraint (e.g., $\Ord~\aA$), which is
constructively reflected in a \SystemF projection function. Thus, our key idea
is to store the instance context within the class dictionary and retrieve it
when necessary using \SystemF projection functions.

This technique poses several technical challenges, which we elaborate on in the
remainder of this section.

\subsection{Challenge 1: Lack of Parametricity}\label{sec:classes-bidirectional-challenges:non-parametric}

Possibly the biggest challenge in interpreting class instances bidirectionally
lies in the non-parametric dictionary representation. To explain what that
means, let us consider the standard equality class $\Eq$
\[
\keyclass~\Eq~\sepia{a}~\keywhere~\{~(\opEq) :: \sepia{a} \to \sepia{a} \to \Bool~\}
\]
along with three instances:
\[
\begin{array}{l@{\hspace{1mm}}c@{\hspace{1mm}}l@{\hspace{1mm}}l}
  \keyinstance                                &             & \Eq~\Int                   & \keywhere~\{~(\opEq) = \dots~\} \\ 
  \keyinstance~\Eq~\sepia{b}                  & \Rightarrow & \Eq~[\sepia{b}]            & \keywhere~\{~(\opEq) = \dots~\} \\ 
  \keyinstance~(\Eq~\sepia{c}, \Eq~\sepia{d}) & \Rightarrow & \Eq~(\sepia{c}, \sepia{d}) & \keywhere~\{~(\opEq) = \dots~\} \\ 
\end{array}
\]
The instance context for each instance varies, depending on the instance
parameter $\aA$.
In the well-established dictionary-passing elaboration
approach~\citep{ClassElaboration}, these contexts correspond
to the following \SystemF types:
\[
\begin{array}{lcl} 
  \aA = \Int      & \implies & \mathit{Ctx}~\aA = \Unit                            \\
  \aA = [\bB]     & \implies & \mathit{Ctx}~\aA = \tyCon_\Eq~\bB                   \\
  \aA = (\cC,\dD) & \implies & \mathit{Ctx}~\aA = (\tyCon_\Eq~\cC, \tyCon_\Eq~\dD) \\
\end{array}
\]
where $\tyCon_\Eq$ is the \SystemF type constructor for the class dictionary.
Obviously, the \SystemF representation of the instance context is not uniform
but varies, depending on how we refine the class parameter $\aA$. This means
that parametricity~\citep{reynolds1983types} as offered by \SystemF is not
sufficient for interpreting instances bidirectionally; a more powerful calculus
is needed as our target language.


\subsection{Challenge 2: Termination of Type Inference}\label{sec:classes-bidirectional-challenges:termination}

The specification of typing is not affected much by bidirectional instances but
this is not the case for type inference. Consider for example the inversion of
the $\Eq~[\bB]$ instance:
\[
\blueforall \bB.~\Eq~[\bB] \Rightarrow \Eq~\bB
\]
If such axioms are not used with care, the termination of the type inference
algorithm is threatened. The standard backwards-chaining entailment
algorithm~\citep{sld-resolution} cannot use such axioms to simplify goals and
terminate. For example, we can ``simplify'' $\Eq~\Int$ to $\Eq~[\Int]$ using
the above axiom. Not only is the size of the constraint bigger than the one we
started with, but the axiom can also be applied infinitely many times (to
capture that all nested list types are instances of $\Eq$): the resolution tree
now contains infinite paths.  Thus, even a backtracking approach (such as the
one used by~\citet{quantcs}) cannot handle bidirectional instances in an
obvious way: bidirectional axioms need to be used selectively to ensure the
termination of type inference.

\subsection{Challenge 3: Principality of Types}\label{sec:classes-bidirectional-challenges:principal-types}

Finally, the introduction of bidirectional instances threatens the principality
of types. In the absence of bidirectional instances, function $\cmp$ has
a single most general type:
\[
  \cmp :: \Eq~\aA \Rightarrow \aA \to \aA \to \Bool
\]
Constraint $\Eq~\aA$ can entail constraint $\Eq~[\aA]$ but not the other way
around.
In a system equipped with bidirectional instances, $\cmp$ can have
multiple most general types. All the following types are equally general:
\[
\begin{array}{ll}
  \cmp :: \Eq~\aA & \Rightarrow \aA \to \aA \to \Bool \\
  \cmp :: \Eq~[[\aA]] & \Rightarrow \aA \to \aA \to \Bool \\
  \cmp :: \Eq~[\Maybe~[\aA]] & \Rightarrow \aA \to \aA \to \Bool \\
\end{array}
\]
This makes specifying the correctness of type inference more difficult, as we
should now infer one type from a set of equally general types.

In vanilla Haskell 98, only the first type annotation is well-formed. By
using the \texttt{FlexibleContexts} extension, all three become well-formed type
annotations, but they are not equivalent: the second and third annotations are
implied by the first, but not the other way around; for the implementation of
$\cmp$ as given at the start of
Section~\ref{sec:classes-bidirectional-challenges} specifically, only the first
annotation will type check.
This annotation is also the only valid (and principal) type.
With bidirectional instances, all three are acceptable types for $\cmp$, and in
fact, all three are \emph{principal types}.

Although it may seem alarming that principal types are not unique in our
extension, this is in fact not new.
The \HM system has the same issue, as well as its extension with qualified
types~\citep{JonesThesis}.
For \HM, principality of types is refined to take into account the
possibilities for positioning universal quantifiers.
Similarly, type classes exhibit the same problem in terms of the order of
constraints, as well as by means of {\em
simplification}~\cite{JonesImprovement}.

In summary, in the presence of bidirectional instances a function can have
infinitely many---equivalent to each other---principal types. This is not
necessarily a problem but in order to ensure well-defined semantics for our
extension, it is imperative that we revisit the notion of {\em type
subsumption}, as well as the definition of the {\em principal type} property.

The next section describes our strategy for dealing with bidirectionality in
intuitive terms; all formal aspects of our extension are described in
Section~\ref{sec:the-basic-system}.

\section{Bidirectional Instances, Informally}\label{sec:classes-bidirectional-formal-extensions}


In this section we describe our approach to interpreting type class instances
bidirectionally, using as an example the elaboration of a variation of function
$\cmp$ (Section~\ref{sec:classes-bidirectional-challenges}):
\[
\begin{array}{c}
  \cmp_2 :: \Eq~(\bB,\cC) \Rightarrow \bB \to \bB \to \cC \to \cC \to \Bool       \\
  \cmp_2~\xX_1~\xX_2~\yY_1~\yY_2 = (\xX_1~\opEq~\xX_2)~\opAnd~(\yY_1~\opEq~\yY_2) \\
\end{array}
\]
Though our formalization targets \SystemFC~\citep{systemfc} (GHC's intermediate
language), we avoid such formality here and we translate type classes to
GHC-flavored Haskell dictionaries instead.

\paragraph{Dictionary Representation}

First, we show how to elaborate declarations. 
For example, we elaborate class $\Eq$
\[
  \keyclass~\Eq~\aA~\keywhere~\{~(\opEq) :: \aA \to \aA \to \Bool~\}
\]
into the following declarations:
\[
\begin{array}{l}
  \keytype~\keyfamily~\tyFam_\Eq~\aA                                                                             \\[2mm]
  \keydata~\tyCon_\Eq~\aA = \dataCon_\Eq~~(\tyFam_\Eq~\aA)~~(\aA \to \aA \to \Bool)                                \\[2mm]
  (\opEq) :: \tyCon_\Eq~\aA \to (\aA \to \aA \to \Bool)                                                \\
  (\opEq)~\dD = \keycase~\dD~\keyof~\{~\dataCon_\Eq~\ctx~\eq \to \eq~\}       \\
  \phantom{\hspace{0.98\columnwidth}} \\[-4mm]
\end{array}
\]
Traditionally, class declarations are elaborated into a dictionary type
($\tyCon_\Eq$) and $n$ functions, each corresponding to a class method. We
extend this approach with an open type function
$\tyFam_\Eq$~\citep{typecheckingwithopentf}, which is meant to capture the
dependency between the class parameter and the instance context. The dictionary
type is extended so that the instance context of type $\tyFam_\Eq~\aA$ is also
stored.

The use of type families (and in general the choice of \SystemFC instead of plain \SystemF)
addresses the challenge of
Section~\ref{sec:classes-bidirectional-challenges:non-parametric}; \SystemFC
offers native support for open, non-parametric type-level functions, which is
exactly what we need, given
\begin{inparaenum}[(a)]
\item
  the non-parametric nature of bidirectionality, and
\item
  the open nature of type classes.
\end{inparaenum}

\paragraph{Inversion Functions}

Particularly interesting is the elaboration of class instances. Take for example
the elaboration of the $\Eq$ instance for tuples
\[
\keyinstance~(\Eq~\bB, \Eq~\cC) \Rightarrow \Eq~(\bB, \cC)~\keywhere~\{~\eq = \ldots~\}
\]
which we elaborate into two kinds of declarations.

The first is a type family instance, mapping the class parameter $(\bB,
\cC)$ to the corresponding instance context representation $(\tyCon_\Eq~\bB,
\tyCon_\Eq~\cC)$:
\[
  \keytype~\keyinstance~\tyFam_\Eq~(\bB, \cC) = (\tyCon_\Eq~\bB, \tyCon_\Eq~\cC)
\]
Its purpose is illustrated below.
%
The next three ($\dD_0$, $\dD_1$, and $\dD_2$) are the {\em dictionary constructors} introduced by the
instance.
The first one---known as the {\em instance axiom}---captures the traditional
meaning of the instance: {\em if $\Eq~\cC$ and $\Eq~\dD$ hold, then so does
$\Eq~(\cC, \dD)$}:
\[
\begin{array}{@{\hspace{0mm}}l@{\hspace{0mm}}}
  \dD_0 :: \tyCon_\Eq~\bB \to \tyCon_\Eq~\cC \to \tyCon_\Eq~(\bB, \cC)                \\
  \dD_0~\dD_b~\dD_c = \dataCon_\Eq~(\dD_b, \dD_c)~(\ldots)                            \\
\end{array}
\]
The next two functions (or, better, dictionary constructors) witness the
inversions of the instance axiom, so we refer to them as the {\em inverted
instance axioms}:
\[
\begin{array}{@{\hspace{0mm}}l@{\hspace{0mm}}}
  \dD_1 :: \tyCon_\Eq~(\bB, \cC) \to \tyCon_\Eq~\bB                                      \\
  \dD_1~(\dataCon_\Eq~\ctx~\xX) = \keycase~\ctx~\keyof~\{~(\dD_b, \dD_c) \to \dD_b~\} \\[2mm]
  \dD_2 :: \tyCon_\Eq~(\bB, \cC) \to \tyCon_\Eq~\cC                                      \\
  \dD_2~(\dataCon_\Eq~\ctx~\xX) = \keycase~\ctx~\keyof~\{~(\dD_b, \dD_c) \to \dD_c~\} \\
\end{array}
\]
$\dD_0$, $\dD_1$, and $\dD_2$ are witnesses of the following
introduction and elimination rules, respectively:
\begin{mathpar}
\inferrule*[right=]
           { \Eq~\cC \\ \Eq~\dD }
           { \Eq~(\cC, \dD) }

\inferrule*[right=]
           { \Eq~(\cC, \dD) }
           { \Eq~\cC }

\inferrule*[right=]
           { \Eq~(\cC, \dD) }
           { \Eq~\dD }
\end{mathpar}
%
The significance of the type family instance also becomes apparent in the
definition of $\dD_0$, $\dD_1$, and $\dD_2$: to store (in the definition of
$\dD_0$) and extract (in the definitions of $\dD_1$ and $\dD_2$) the instance
context, we need to change its type from $(\tyCon_\Eq~\bB, \tyCon_\Eq~\cC)$ to
$(\tyFam_\Eq~\aA)$, and vice versa. In source-level Haskell such conversions
are implicit (like in the code above), but in \SystemFC, they are explicit (see
Section~\ref{sec:the-basic-system:target-syntax}). Our elaboration algorithm
(Section~\ref{sec:the-basic-system}) explains this translation in detail.


Finally, it is worth mentioning that the example already illustrates one of our
design choices: instead of directly introducing a logical biconditional
connective into our calculus, we simplify matters by generating the inversions
as separate functions.  This allows us to reuse existing infrastructure and the
well-established dictionary-passing elaboration method.

\paragraph{Additional Derivations}

Finally, function $\cmp_2$ is elaborated as follows:
\[
\begin{array}{l@{\hspace{1mm}}c@{\hspace{1mm}}l}
  \multicolumn{3}{l}{\cmp_2 :: \tyCon_\Eq~(\bB,\cC) \to \bB \to \bB \to \cC \to \cC \to \Bool} \\
  \cmp_2~\dD~\xX_1~\xX_2~\yY_1~\yY_2 & = & \keylet~\dD_1' : \tyCon_\Eq~\bB = \dD_1~\dD~\keyin \\
                                     &   & \keylet~\dD_2' : \tyCon_\Eq~\cC = \dD_2~\dD~\keyin \\
                                     &   & ((\opEq)~\dD_1'~\xX_1~\xX_2)~\opAnd~((\opEq)~\dD_2'~\yY_1~\yY_2) \\
\end{array}
\]
The implementation of $\cmp_2$ requires $\Eq~\bB$ and
$\Eq~\cC$, but the signature provides $\Eq~(\bB,\cC)$. This is remedied by
using the dictionary constructors $\dD_1$ and $\dD_2$ defined above to locally
extract the needed information from the given dictionary $\dD$. As we
illustrate below (Section~\ref{subsubsec:transitive-closure}), the creation of
such a context might need several steps, but is guaranteed to terminate if the
instances respect well-established termination conditions
(Section~\ref{sec:classes-bidirectional-metatheory:termination}). Hence, our
approach also addresses the challenge described in
Section~\ref{sec:classes-bidirectional-challenges:termination}.


\section{Type Classes with Superclasses}
\label{sec:basic-with-superclasses}

Before we can present our formalization of bidirectional instances in
Section~\ref{sec:the-basic-system}, in this section we present a formalization
of type classes with superclasses, including the specification of typing and
elaboration to \SystemFC, as well as a type inference and elaboration
algorithm.

This detour serves two purposes. Firstly, our extension reuses the
infrastructure used by superclasses, so introducing superclasses first allows
us to focus on the feature-specific changes alone in the next section.
Secondly, to our knowledge, we are the first to formalize type inference and
elaboration of type classes in the presence of superclass constraints, so this
section is a contribution in its own right (in particular
Section~\ref{sec:the-basic-system:algorithms}).


The presentation of this section
is deliberately technical, as it is aimed
to serve as a specification for verification and implementation of our feature.
Indeed, our prototype, which can be found at
\url{https://github.com/KoenP/bidirectional-instances},
follows our specification closely.

The remainder of this section is structured as follows:
Section~\ref{sec:the-basic-system:source-syntax} presents the syntax of source
programs and Section~\ref{sec:the-basic-system:target-syntax} gives the syntax
of \SystemFC. The specification of typing and elaboration is given in
Section~\ref{sec:the-basic-system:specifications}, while
Section~\ref{sec:the-basic-system:algorithms} presents a type inference with
elaboration algorithm. To simplify the presentation, throughout the whole
section we highlight the parts of the rules that relate to elaboration.

\paragraph{A note on notation.}
From this section onwards, we will use overline notation to denote indexed sequences.
For instance, when we write $\overline{x}^n$, we mean a sequence $x_1, x_2,
..., x_n$. Sometimes we omit the multiplicity superscript if the number of
elements is of no interest (so $\overline{x}$ means $x_1, x_2, ..., x_n$ for
some $n$).  In some cases we use the overline notation on more complex
structures than just variables, if we believe the meaning is clear from context
(for instance, we might write $\lambda~\overline{(x : d)}^n.~t$ to mean
$\lambda~(x_1 : d_1)~(x_2 : d_2)~\ldots~(x_n : d_n).~t$).

\subsection{Source Syntax}\label{sec:the-basic-system:source-syntax}

\begin{figure}
\small
    \centering
    \begin{subfigure}[b]{\columnwidth}
        \[
        \begin{array}{@{\hspace{0mm}}c@{\hspace{0mm}}}
        \begin{tabular*}{0.38\columnwidth}{@{\hspace{0mm}}l@{\hspace{1mm}}c@{\hspace{1mm}}l@{\extracolsep{\fill}}r@{\hspace{0mm}}}
          \syntaxLine{\program}
                     {::=}
                     {\overline{\decl}}
                     {} 
          \\
          \syntaxLine{\decl}
                     {::=}
                     {\clsDecl \mid \insDecl \mid \valDecl}
                     {} 
          \\
          \syntaxLine{}{}{}{}
        \end{tabular*}
        ~ 
        \begin{tabular*}{0.3\columnwidth}{@{\hspace{0mm}}l@{\hspace{1mm}}c@{\hspace{1mm}}l@{\extracolsep{\fill}}r@{\hspace{0mm}}}
          \syntaxLine{\monotype}
                     {::=}
                     {a \mid \monotype_1 \to \monotype_2}
                     {} 
          \\
          \syntaxLine{\qualtype}
                     {::=}
                     {\monotype \mid \constraint \Rightarrow \qualtype}
                     {} 
          \\
          \syntaxLine{\polytype}
                     {::=}
                     {\qualtype \mid \blueforall a.~\polytype}
                     {} 
        \end{tabular*}
        ~ 
        \begin{tabular*}{0.3\columnwidth}{@{\hspace{0mm}}l@{\hspace{1mm}}c@{\hspace{1mm}}l@{\extracolsep{\fill}}r@{\hspace{0mm}}}
          \syntaxLine{\constraintScheme}
                     {::=}
                     {\blueforall \as.~\constraintSet \Rightarrow \constraint}
                     {} 
          \\
          \syntaxLine{\constraintSet}
                     {::=}
                     {\bullet \mid \constraintSet, \constraint}
                     {} 
          \\
          \syntaxLine{\constraint}
                     {::=}
                     {\TC~\monotype}
                     {} 
        \end{tabular*}
        \\
        \begin{tabular*}{\columnwidth}{@{\hspace{0mm}}l@{\hspace{1mm}}c@{\hspace{1mm}}l@{\extracolsep{\fill}}r@{\hspace{0mm}}}
          \syntaxLine{e}
                     {::=}
                     {x \mid \lambda x.~e \mid e_1~e_2 \mid \keylet~x = e_1~\keyin~e_2}
                     {} 
        \end{tabular*}
        \\ 
        \begin{tabular*}{\columnwidth}{@{\hspace{0mm}}l@{\hspace{1mm}}c@{\hspace{1mm}}l@{\extracolsep{\fill}}r@{\hspace{0mm}}}
          \syntaxLine{\clsDecl}
                     {::=}
                     {\keyclass~\blueforall a.~\constraintSet \Rightarrow \TC~a~\keywhere~\{~f :: \polytype~\}}
                     {} 
          \\
          \syntaxLine{\insDecl}
                     {::=}
                     {\keyinstance~\blueforall \bs.~\constraintSet \Rightarrow \TC~\monotype~\keywhere~\{~f = e~\}}
                     {} 
          \\
          \syntaxLine{\valDecl}
                     {::=}
                     {x = e \mid x :: \polytype = e}
                     {} 
        \end{tabular*}
        \end{array}
        \]
        \vspace{-3mm}
        \caption{Basic System: Syntax}
        \label{subfig:class-background:tc-syntax}
    \end{subfigure}

    \begin{subfigure}[b]{\columnwidth}
        \[
        \begin{array}{@{\hspace{0mm}}c@{\hspace{0mm}}}
        \begin{tabular*}{\columnwidth}{@{\hspace{0mm}}l@{\hspace{2mm}}c@{\hspace{1mm}}l@{\extracolsep{\fill}}r@{\hspace{0mm}}}
          \syntaxLine{\fcType}
                     {::=}
                     {a \mid \tyCon \mid \fcType_1~\fcType_2 \mid \blueforall a.~\fcType \mid \tyFamApp{\tyFam}{\fcTypes}
                      \mid \fcEqCt \Rightarrow \fcType}
                     {} 
          \\
          \syntaxLine{\tyPat}
                     {::=}
                     {a \mid \tyCon \mid \tyPat_1~\tyPat_2}
                     {} 
        \end{tabular*}
        \\ 
        \begin{tabular*}{\columnwidth}{@{\hspace{0mm}}l@{\hspace{2mm}}c@{\hspace{1mm}}l@{\extracolsep{\fill}}r@{\hspace{0mm}}}
          \syntaxLine{\fcEqCt}
                     {::=}
                     {\fcType_1 \sim \fcType_2}
                     {} 
        \end{tabular*}
        \\ 
        \begin{tabular*}{\columnwidth}{@{\hspace{0mm}}l@{\hspace{2mm}}c@{\hspace{1mm}}l@{\extracolsep{\fill}}r@{\hspace{0mm}}}
          \syntaxLine{\coercion}
                     {::=}
                     {\frefl{\fcType}                     \mid
                      \fsym{\coercion}                    \mid
                      \fleft{\coercion}                   \mid
                      \fright{\coercion}                  \mid
                      \fcomp{\coercion_1}{\coercion_2}    \mid
                      \fcEqCt \Rightarrow \coercion}
                     {} 
          \\
          \syntaxLine{  }
                     {\mid}
                     {\tyFamApp{\tyFam}{\coercions}        \mid
                      \blueforall a.~\coercion                 \mid
                      \finst{\coercion_1}{\coercion_2}     \mid
                      g~\fcTypes                           \mid
                      \covar                               \mid
                      \finstQual{\coercion_1}{\coercion_2} \mid
                      \coercion_1~\coercion_2}
                     {  }
        \end{tabular*}
        \\ 
        \begin{tabular*}{\columnwidth}{@{\hspace{0mm}}l@{\hspace{2mm}}c@{\hspace{1mm}}l@{\extracolsep{\fill}}r@{\hspace{0mm}}}
          \syntaxLine{t}
                     {::=}
                     {x                                          \mid
                      \dataCon                                   \mid
                      \Lambda a.~t                               \mid
                      t~\fcType                                  \mid
                      \lambda (x : \fcType).~t                   \mid
                      t_1~t_2                                    \mid
                      \Lambda (\covar : \fcEqCt).~t}
                     {} 
          \\
          \syntaxLine{ }
                     {\mid}
                     {t~\coercion                                   \mid
                      \fcCast{t}{\coercion}                         \mid
                      \keycase~t_1~\keyof~\overline{\fcPat \to t_2} \mid
                      \keylet~x : \fcType = t_1~\keyin~t_2}
                     { }
        \end{tabular*}
        \\ 
        \begin{tabular*}{\columnwidth}{@{\hspace{0mm}}l@{\hspace{2mm}}c@{\hspace{1mm}}l@{\extracolsep{\fill}}r@{\hspace{0mm}}}
          \syntaxLine{\fcPat}
                     {::=}
                     {\dataCon~\bs~(\overline{\covar : \fcEqCt})~(\overline{x : \fcType})}
                     {} 
        \end{tabular*}
        \\ 
        \begin{tabular*}{\columnwidth}{@{\hspace{0mm}}l@{\hspace{2mm}}c@{\hspace{1mm}}l@{\extracolsep{\fill}}r@{\hspace{0mm}}}
          \syntaxLine{\decl}
                     {::=}
                     {\keydata~\tyCon~\as~\keywhere~\{~\overline{\dataCon : \fcType}~\}
                      \mid \keytype~\tyFamApp{\tyFam}{\as}
                     }
                     {} 
          \\
          \syntaxLine{ }
                     {\mid}
                     {\keyaxiom~g~\as : \tyFamApp{\tyFam}{\tyPats} \sim \fcType
                      \mid \keylet~x : \fcType = t
                     }
                     {} 
        \end{tabular*}
        \end{array}
        \]
        \vspace{-3mm}
        \caption{\SystemFC: Syntax}
        \label{subfig:class-background:fc-syntax}
    \end{subfigure}
    \vspace{-7mm}
    \caption{Source and Target Syntax}
    \label{fig:all-syntax}
\end{figure}

The syntax of the basic system is presented in
Figure~\ref{subfig:class-background:tc-syntax}.
A program $\program$ consists of a list of declarations $\decl$, which can be
class declarations $\clsDecl$, instance declarations $\insDecl$, or value
bindings $\valDecl$.
The syntax of class declarations, instances, and value bindings is standard. In
order to reduce the notational burden, we omit all mention of kinds and assume
that each class has exactly one method\footnote{Adding multiple methods would
only increase verbosity without significant gains, since we would only have to
add (many) overbars to the typing rules.}. Additionally, we explicitly quantify over
the type variables $\as$ that are bound in the class/instance head and context.
Expressions comprise a $\lambda$-calculus, extended with let bindings.
Types are stratified in monotypes $\monotype$, qualified types $\qualtype$, and
polytypes $\polytype$. This is standard practice for \HM extended with
qualified types~\citep{qualtypes}.
Next, the syntax of constraints is straightforward: constraint schemes
$\constraintScheme$ capture implications generated by class and instance
declarations. Sets of constraints (like superclass constraints or instance
contexts) are denoted by $\constraintSet$ and single class constraints are
denoted by $\constraint$.

\subsection{Target Syntax}\label{sec:the-basic-system:target-syntax}

The syntax of \SystemFC programs is presented in
Figure~\ref{subfig:class-background:fc-syntax}. In contrast to prior
specifications of type classes that use \SystemF as the target language for
elaboration, our elaboration targets \SystemFC. Though for plain type classes
this is not required (\SystemF is a strict subset of \SystemFC), it is
essential for bidirectional instances, as we explained in
Section~\ref{sec:classes-bidirectional-challenges:non-parametric}.

Types $\fcType$ include all \SystemF types, extended with type family
applications $\tyFamApp{\tyFam}{\fcTypes}$, and qualified types $(\fcEqCt
\Rightarrow \fcType)$. Qualified types classify terms that use coercion
abstraction. 
Type patterns $\tyPat$ are---as expected---the predicative subset of types.

Next, Figure~\ref{subfig:class-background:fc-syntax} presents proposition types
$\fcEqCt$, capturing equalities between types. In the same way that types classify terms,
propositions classify coercions $\coercion$; a coercion is nothing more than a
proof of type equality and can take any of the following forms:

Reflexivity $\frefl{\fcType}$, symmetry $(\fsym{\coercion})$ and transitivity
$(\fcomp{\coercion_1}{\coercion_2})$ express that type equality is an
equivalence relation. Forms $\tyFamApp{\tyFam}{\coercions}$ and
$(\coercion_1~\coercion_2)$ capture injection, while $(\fleft{\coercion})$ and
$(\fright{\coercion})$ capture projection, which follows from the injectivity
of type application.
Equality for universally quantified and qualified types is witnessed by forms
$\blueforall a.~\coercion$ and $\fcEqCt \Rightarrow \coercion$, respectively.
Similarly, forms $\finst{\coercion_1}{\coercion_2}$ and
$\finstQual{\coercion_1}{\coercion_2}$ witness the equality of type
instantiation or coercion application, respectively.

Additionally, \SystemFC introduces two new symbol classes: coercion variables
$\covar$ and axiom names $g$.
The former represent local constraints and are introduced by explicit coercion
abstraction or GADT pattern matching. The latter constitute the axiomatic part
of the theory, and are generated from top-level axioms, which correspond to
type family instances or newtype declarations~\citep{haskell98}. As we
illustrated in the previous section, our bidirectional interpretation of class
instances also gives rises to such axioms.

The semantics of the coercion forms we gave above is formally captured in
coercion typing $\fcTcCo{\tyEnv}{\coercion}{\fcEqCt}$, which we include in
Appendix~\ref{appendix:remaining-systemfc}.

\SystemFC terms $t$ also conservatively extend \SystemF terms. The
interesting new forms are coercion abstraction $(\Lambda (\covar : \fcEqCt).~t)$,
coercion application $(t~\coercion)$, and type casts $(\fcCast{t}{\coercion})$.
In simple terms, if $\coercion$ is a proof that $\fcType_1$ is equal to
$\fcType_2$ and $t$ has type $\fcType_1$, then $(\fcCast{t}{\coercion})$ has
type $\fcType_2$.
Patterns $\fcPat$ capture existential variables $\bs$ and local equality
constraints $(\overline{\covar : \fcEqCt})$ in addition to term variables
$\xs$, to account for GADTs.

Programs consist of declarations $\decl$, which consist of datatype
declarations, type family declarations, type equality axioms, and variable
bindings.

\subsection{Additional Constructs}\label{sec:the-basic-system:additional-constructs}

In order to state the specification of typing and elaboration succinctly, we
first introduce some additional notation.
First, we introduce typing environments and program theories:
\[\small
\begin{array}{@{\hspace{0mm}}c@{\hspace{0mm}}}
\begin{tabular*}{0.9\columnwidth}{@{\hspace{0mm}}l@{\hspace{1mm}}c@{\hspace{1mm}}l@{\extracolsep{\fill}}r@{\hspace{0mm}}}
  \syntaxLine{\tyEnv}
             {::=}
             {\bullet \mid \tyEnv, a \mid \tyEnv, x : \polytype}
             {typing environment}
  \\
  \syntaxLine{\programTheory}
             {::=}
             {\langle \superAxiomSet, \instanceAxiomSet, \localTheory \rangle}
             {program theory}
\end{tabular*}
\end{array}
\]
Typing environments are standard. The program theory $\programTheory$ contains
schemes generated by class and instance declarations, and gets
extended with {\em local constraints}, when going under a qualified type. We
explicitly represent the program theory as a triple of
the superclass axioms $\superAxiomSet$, the instance axioms
$\instanceAxiomSet$, and the local axioms $\localTheory$. We use the notation
$\programTheory\lComma d : \constraint$ to denote that we extend the local
component of the triple, and similar notation for the other components.

Note that the specification we present below treats $\programTheory$ as a
conflated constraint set (that is, if
$P = \langle \superAxiomSet, \instanceAxiomSet, \localTheory \rangle$ we write
$(d:\constraintScheme) \in \programTheory$
to mean
$(d:\constraintScheme) \in \superAxiomSet \cup \instanceAxiomSet \cup \localTheory$),
while the inference algorithm we present in
Section~\ref{sec:the-basic-system:algorithms} distinguishes between the
subsets; such a formalization is closer to actual implementations of type
classes.

Finally, we define evidence-annotated axiom sets $\axiomSet$, local axioms
$\annConstraintSet$, and constraints $\annConstraint$:
\[
\begin{array}{@{\hspace{0mm}}c@{\hspace{0mm}}}
\begin{tabular*}{0.9\columnwidth}{@{\hspace{0mm}}l@{\hspace{1mm}}c@{\hspace{1mm}}l@{\extracolsep{\fill}}r@{\hspace{0mm}}}
  \syntaxLine{\axiomSet}
             {::=}
             {\bullet \mid \axiomSet, \highlight{d :} \constraintScheme}
             {variable-annotated axiom set}
  \\
  \syntaxLine{\annConstraintSet}
             {::=}
             {\bullet \mid \annConstraintSet, \highlight{d :} \constraint}
             {variable-annotated constraint set}
  \\
  \syntaxLine{\annConstraint}
             {::=}
             {\highlight{d :} \constraint}
             {variable-annotated class constraint}
\end{tabular*}
\end{array}
\]
This notation allows us to present typing and elaboration succinctly below.

\subsection{Specification of Typing and Elaboration}\label{sec:the-basic-system:specifications}

\begin{figure*}
\small
\begin{minipage}{\textwidth}
\begin{flushleft}
\namedRuleform{\tcTcClsDecl{\tyEnv}{\clsDecl}{\programTheory_S}{\tyEnv_c}{\overline{\decl}}}
              {Class Declaration Typing}
\end{flushleft}
\begin{mathpar}
\inferrule*[right=Cls]
           { \metaforall{\constraint_i \in \constraints^n}
               \tcTcCt{\tyEnv, a}{\constraint_i}{\fcType_i} \\
             \tcTcTy{\tyEnv, a}{\polytype}{\fcType} \\ 
             \programTheory_S = [\metaforall{\constraint_i \in \constraints^n}\highlight{d_i :}~~ \blueforall a.~\TC~a \Rightarrow \constraint_i] \\
             \tyEnv_c = [f : \blueforall a.~\TC~a \Rightarrow \polytype] \\\\ 
             \highlight{
               \decl_M = ~\keylet~f : \blueforall a.~\tyCon_\TC~a \to \fcType
               = \Lambda a.~\lambda (d : \tyCon_\TC~a).~\genProject{n+1}{\TC}{d}
             }
             \\
             \highlight{
               \decls_d =
               \metaforall{i \in [1..n]} \keylet~d_i : \blueforall a.~\tyCon_\TC~a \to \fcType_i = \Lambda a.~\lambda (d : \tyCon_\TC~a).~\genProject{i}{\TC}{d}
             }
           } 
           { \tcTcClsDecl{\tyEnv}
                         {\keyclass~\blueforall a.~\constraints^n \Rightarrow \TC~a~\keywhere~\{~f :: \polytype~\}}
                         {\programTheory_S}
                         {\tyEnv_c}
                         {[\keydata~\tyCon_\TC~a = \dataCon_\TC~\fcTypes~\fcType,
                            \decl_M, \decls_d]}
           }
\end{mathpar}

\begin{flushleft}
\namedRuleform{\tcTcInsDecl{\programTheory}{\tyEnv}{\insDecl}{\programTheory_i}{\decl}}
              {Class Instance Typing}
\end{flushleft}
\begin{mathpar}
\inferrule*[right=Ins]
           {
             \keyclass~\blueforall a.~\supconstraints^n \Rightarrow \TC~a~\keywhere~\{~f :: \polytype~\} \\ 
             \tyEnv_I = \tyEnv, \bs \\
             \programTheory_I = \programTheory\lComma
               \metaforall{\constraint_i \in \constraints^m}\highlight{d_i :} \constraint_i \\
             \metaforall{\constraint_i \in \constraints^m}
               \tcTcCt{\tyEnv_I}{\constraint_i}{\fcType_i} \\
             \tcTcTy{\tyEnv_I}{\monotype}{\fcType_0} \\
             \constraintScheme = \blueforall \bs.~\overline{\constraint}^m \Rightarrow \TC~\monotype \\
             \metaforall{\supconstraint_i \in \supconstraints^n}
               \tcEntailCt{\programTheory_I}{\tyEnv_I}{t_i}{[\monotype/a]\supconstraint_i} \\
             \tcTcTm{\programTheory_I\iComma \highlight{d :} \constraintScheme}{\tyEnv_I}{e}{[\monotype/a]\polytype}{t} \\
           } 
           { \tcTcInsDecl{\programTheory}
                         {\tyEnv}
                         {\keyinstance~\blueforall \bs.~\overline{\constraint}^m \Rightarrow \TC~\monotype~\keywhere~\{~f = e~\}}
                         {[\highlight{d :} \constraintScheme]}
                         {\keylet~d : \blueforall \bs.~\fcTypes^m \to \tyCon_\TC~\fcType_0 = \Lambda \bs.~\lambda (\overline{d : \fcType}^m).~\dataCon_\TC~\fcType~\ts^n~t}
           }
\end{mathpar}

\begin{flushleft}
\namedRuleform{\tcTcValDecl{\programTheory}{\tyEnv}{\valDecl}{\tyEnv'}{\decl}}
              {Value Binding Typing}
\end{flushleft}
\begin{mathpar}
\inferrule*[right=Sig]
           { \tcTcTy{\tyEnv}{\polytype}{\fcType} \\
             \tcTcTm{\programTheory}{\tyEnv, x : \polytype}{e}{\polytype}{t} 
           } 
           { \tcTcValDecl{\programTheory}
                         {\tyEnv}
                         {x :: \polytype = e}
                         {[x : \polytype]}
                         {\keylet~x : \fcType = t}
           }

\inferrule*[right=Val]
           { \metaforall{\constraint_i \in \constraints}
               \tcTcCt {\tyEnv,\as} {\constraint_i} {\fcType_i} \\
             \tcTcCt{\tyEnv, \as}{\monotype}{\fcType} \\
             \tcTcTm{\programTheory, \overline{\highlight{d :} \constraint}}{\tyEnv, \as, x : \monotype}{e}{\monotype}{t} \\\\
             \highlight{\decl = \keylet~x : \blueforall \as.~\fcTypes \to \fcType = \Lambda \as.~\lambda (\overline{d : \fcType}).~[x~\as~\ds/x]t}
           } 
           { \tcTcValDecl{\programTheory}
                         {\tyEnv}
                         {x = e}
                         {[x : \blueforall \as.~\overline{\constraint} \Rightarrow \monotype]}
                         {\decl}
           }
\end{mathpar}
\end{minipage}
\vspace{-3mm}
\caption{Basic System: Declaration Typing and Elaboration into \SystemFC}
\label{fig:basic-system-specification}
\end{figure*}

\subsubsection{Term, Type, and Constraint Typing}
\label{subsubsec:basic-system:term-type-constraint-spec}

Since most of the specification of typing for our core calculus can be found in
prior work (see for example the work of~\citet{fundeps-core}), we omit the
definitions for term typing, type well-formedness, and constraint
well-formedness from our main presentation; they can be found in
Appendix~\ref{appendix:remaining-specification}. Their signatures are the
following:
\[
\begin{array}{@{\hspace{0mm}}c@{\hspace{0mm}}}
\begin{tabular*}{0.9\columnwidth}{@{\hspace{0mm}}l@{\extracolsep{\fill}}r@{\hspace{0mm}}}
  $\tcTcTy{\tyEnv}{\polytype}{\fcType}$              & \textit{type well-formedness}       \\
  $\tcTcCt{\tyEnv}{\constraint}{\fcType}$            & \textit{constraint well-formedness} \\
  $\tcTcTm{\programTheory}{\tyEnv}{e}{\polytype}{t}$ & \textit{term typing}                \\
\end{tabular*}
\end{array}
\]
Type well-formedness ensures that $\polytype$ is
well-formed under typing environment $\tyEnv$ and elaborates it into \SystemFC type
$\fcType$.
Constraint well-formedness ensures that the
type class constraint $\constraint$ is
well-formed under typing environment $\tyEnv$ and elaborates it into \SystemFC
dictionary type $\fcType$.
The term typing relation ensures that
$e$ has type $\polytype$ under typing environment $\tyEnv$ and
program theory $\programTheory$, and elaborates $e$ into \SystemFC term $t$.

We focus here on the more relevant aspects of the specification:
constraint entailment and declaration typing.

\subsubsection{Constraint Entailment}
\label{subsubsec:basic-system:entailment-spec}

The notion of constraint entailment refers to the resolution of {\em wanted}
constraints, arising from calling overloaded functions, using {\em given}
constraints, provided by type signatures or GADT pattern
matching~\citep{outsideinx}. This procedure is captured by relation
$\tcEntailCtNoShade{\programTheory}{\tyEnv}{t}{\constraint}$, read as
{\em``under given constraints $\programTheory$ and typing environment $\tyEnv$,
  \SystemFC term $t$ is a proof for constraint $\constraint$''}.
\footnote{
  \label{fn:highlights}
  To aid readability, we highlight all aspects of the rules that are concerned
  with elaboration.
}
It is given by a
single rule:
\begin{mathpar}
\small
\inferrule*[right=]
           {(d : \blueforall \as^n.~\constraints^m \Rightarrow \TC~\monotype) \in \programTheory \\
            \metaforall{\monotype_i \in \monotypes^n}
              {\tcTcTy{\tyEnv}{\monotype_i}{\fcType_i}} \\
            \metaforall{\constraint_i \in \constraints}
              \tcEntailCt{\programTheory}{\tyEnv}{t_i}{[\monotypes/\as]\constraint_i}
           } 
           {\tcEntailCt{\programTheory}{\tyEnv}{d~\fcTypes^n~\ts^m}{[\monotypes^n/\as^n](\TC~\monotype)}}
\end{mathpar}
This method of entailment---known as {\em Selective Linear Definite (SLD)
clause resolution}~\citep{sld-resolution} or {\em backwards chaining}---is
the standard sound and complete resolution for Horn clauses.
%
Essentially, we match the head of a given Horn clause in the program theory
$\programTheory$ with the goal, and recursively resolve the premises of the
clause. Dictionary construction behaves accordingly: the selected dictionary
transformer $d$ is instantiated appropriately (i.e., applied to types
$\fcTypes^n$), and then applied to the proofs for the premises, $\ts^m$.

\subsubsection{Declaration Typing}
\label{subsubsec:basic-system:declaration-spec}

The specification of typing with elaboration of declarations is presented in
Figure~\ref{fig:basic-system-specification}.\footnoteref{fn:highlights}
We do not clutter
the rules with freshness conditions by adopting the~\citet{barendregt}
convention.

\paragraph{Class Declarations}

Judgment
$\tcTcClsDeclNoShade{\tyEnv}{\clsDecl}{\programTheory_S}{\tyEnv_c}{\overline{\decl}}$
handles class declarations and is given by Rule~\textsc{Cls}. Apart from
checking the well-scopedness of the class context and the method signature, it
also gives rise to typing environment extension $\tyEnv_c$ which captures the
method type, and program theory extension $\programTheory_S$ which captures the
superclass axioms. All this information is also captured in the generated
declarations $\overline{\decl}$, which includes the dictionary type declaration
$\tyCon_\TC$, the superclass axioms $\ds^n$, and the method $f$. We use
$\genProject{i}{\TC}{d}$ to denote the extraction of the $i$-th field of a
class dictionary $d$ of type $\tyCon_\TC~a$:
\[
\genProject{i}{\TC}{d} \equiv \keycase~d~\keyof~\{~\dataCon_\TC~\xs \to x_i~\}
\]

\paragraph{Instance Declarations}

Judgment
$\tcTcInsDeclNoShade{\programTheory}{\tyEnv}{\insDecl}{\programTheory_i}{\decl}$
handles instance declarations and is also given by a single rule.
Rule~\textsc{Ins} is for the most part straightforward: we ensure that all
objects are well-scoped, and additionally check
\begin{inparaenum}[(a)]
\item
  the entailment of superclass constraints
  (denoted by $\supconstraints^n$)
  and
\item
  the method implementation against its expected type.
\end{inparaenum}
Lastly, the program theory extension induced by the instance is captured in the
unidirectional scheme $\constraintScheme$, which is also elaborated into
\SystemFC dictionary transformer $d$.

\paragraph{Value Bindings}

In order to showcase the usability of bidirectional instances, we have included
in our source language both annotation-free and type-annotated value bindings.

Rule~\textsc{Val} deals with the former, while Rule~\textsc{Sig} deals with the
latter. Since type inference is undecidable in the presence of polymorphic
recursion without type-annotations~\citep{UndecidablePolymorphicRecursion},
Rule~\textsc{Val} ensures that $x$ is used monomorphically in recursive
positions.\footnote{Like all Haskell systems, our specification and
inference algorithm allow polymorphic recursion in the presence of explicit
type annotations (i.e., in class method implementations and
annotated top-level bindings).} Apart from that, both rules are
straightforward.

\subsection{Type Inference with Elaboration}\label{sec:the-basic-system:algorithms}

Now, we give an algorithm for type inference with elaboration. As is standard
practice for HM-based systems, the algorithm proceeds in two phases: constraint
generation and constraint solving.

\subsubsection{Intermediate Constructs}

First, we introduce three intermediate constructs: sets of equality constraints
$\equalities$, type substitutions $\tySubst$, and evidence substitutions
$\evSubst$:
\[
\begin{array}{@{\hspace{0mm}}c@{\hspace{0mm}}}
\begin{tabular*}{0.9\columnwidth}{@{\hspace{0mm}}l@{\hspace{2mm}}c@{\hspace{2mm}}l@{\extracolsep{\fill}}r@{\hspace{0mm}}}
  \syntaxLine{\equalities}
             {::=}
             {\bullet \mid \equalities, \monotype_1 \sim \monotype_2}
             {type equalities}
  \\
  \syntaxLine{\tySubst}
             {::=}
             {\bullet \mid \tySubst \cdot [\monotype/a]}
             {type substitution}
  \\
  \syntaxLine{\evSubst}
             {::=}
             {\bullet \mid \evSubst \cdot [t/d]}
             {evidence substitution}
\end{tabular*}
\end{array}
\]
Type equalities $\equalities$ are generated from the source text (alongside
wanted class constraints $\annConstraintSet$). Type and evidence substitutions
are the results of constraint solving: the former maps unification variables to
types, and the latter maps dictionary variables to dictionaries.

\subsubsection{Elaboration of Terms, Types, and Constraints}

Elaboration of terms, types, and constraints for our core calculus is also
standard and can be found in prior work (see for example the work of~\citet{quantcs}).
The signatures of the corresponding judgments are the following:
\[
\begin{array}{@{\hspace{0mm}}c@{\hspace{0mm}}}
\begin{tabular*}{0.9\columnwidth}{@{\hspace{0mm}}l@{\extracolsep{\fill}}r@{\hspace{0mm}}}
  $\tcElabTmNoShade{\tyEnv}{e}{\monotype}{t}{\annConstraintSet}{\equalities}$         & \textit{term elaboration}       \\
  $\tcElabTy{\polytype} = \fcType$                                                    & \textit{type elaboration}       \\
  $\tcElabCt{\constraint} = \fcType$                                                  & \textit{constraint elaboration} \\
\end{tabular*}
\end{array}
\]
Given a typing environment $\tyEnv$ and a source expression $e$, constraint
generation infers a monotype $\monotype$ for $e$ and generates wanted
constraints $\annConstraintSet$ and $\equalities$, while at the same time
elaborates $e$ into a \SystemFC term $t$.

Elaboration of types and constraints is straightforward: the former elaborates
a source type $\polytype$ into a \SystemFC type $\fcType$, and the latter
transforms a class constraint $\constraint$ into a \SystemFC (dictionary) type
$\fcType$.

Since all three are straightforward, we omit their definition; they can be
found in Appendix~\ref{appendix:remaining-specification}.

\subsubsection{Constraint Solving}

The type class and equality constraints derived from terms are solved with the
following two algorithms.

\paragraph{Solving Equality Constraints}

We solve a set of equality constraints $\equalities$ using the standard
first-order unification algorithm~\cite{DamasMilner}.
Function $\tcUnify{\as}{\equalities} = \tySubst_\bot$ takes a set of equalities
$\equalities$ and, if successful, produces as a result a type substitution
$\tySubst$. The additional argument $\as$ captures the ``untouchable''
variables introduced by type signatures, that is, variables that cannot be
substituted (they can be unified with themselves though). Since its definition
is straightforward, we omit its formal definition; it can be found in
Appendix~\ref{appendix:remaining-specification}.

\paragraph{Solving Class Constraints}

The judgment for solving class constraints takes the form
%
$\tcSimplifyAllCtNoShade{\as}{\axiomSet}{\annConstraintSet_1}{\annConstraintSet_2}{\evSubst}$
and is given by the following rules:
\vspace{-4mm}
\begin{mathpar}
\small
\inferrule*[right=]
           { \nexists \annConstraint \in \annConstraintSet : \tcSimplifyAllCtNoShade{\as}{\axiomSet}{\annConstraint}{\annConstraintSet'}{\evSubst} }
           { \tcSimplifyAllCt{\as}{\axiomSet}{\annConstraintSet}{\annConstraintSet}{\bullet} }
\quad
\inferrule*[right=]
           {\tcSimplifyCt{\as}{\axiomSet}{\annConstraint}{\annConstraintSet_1}{\evSubst_1} \\\\
            \tcSimplifyAllCt{\as}{\axiomSet}{\annConstraintSet, \annConstraintSet_1}{\annConstraintSet_2}{\evSubst_2}
           } 
           {\tcSimplifyAllCt{\as}{\axiomSet}{\annConstraintSet, \annConstraint}{\annConstraintSet_2}{(\evSubst_2 \cdot \evSubst_1)}}
\end{mathpar}
Given a set of untouchable type variables $\as$ and an axiom set $\axiomSet$,
it (exhaustively) replaces a set of constraints $\annConstraintSet_1$ with a
set of {\em simpler}  constraints $\annConstraintSet_2$. This simplification it
achieves via judgment
$\tcSimplifyCtNoShade{\as}{\axiomSet}{\annConstraint}{\annConstraintSet}{\evSubst}$,
given by a single rule:
\begin{mathpar}
\small
\inferrule*[right=Ent]
           {(\highlight{d_I :} \blueforall \bs.~\overline{\constraint}^n \Rightarrow \TC~\monotype_2) \in \axiomSet \\
            \tcUnify{\as}{\monotype_1 \sim \monotype_2} = \tySubst \\
            \annConstraintSet = \metaforall{\constraint_i \in \constraints^n}
               {\highlight{d_i :} \tySubst(\constraint_i)}
           } 
           {\tcSimplifyCt {\as} {\axiomSet}
             {\highlight{d :} \TC~\monotype_1}
             {\annConstraintSet}
             {[d_I~\tySubst(\bs)~\ds^n/d]}}
\end{mathpar}
This form differs from the specification we gave in
Section~\ref{subsubsec:basic-system:entailment-spec} in three ways.

First, we allow constraints to be partially entailed, which allows for {\em
simplification}~\citep{JonesImprovement} of top-level signatures. This is
standard practice in Haskell when inferring types. For instance, when inferring
the signature for
$(\fF~\xX = [\xX]~\opEq~[\xX])$.
Haskell simplifies the derived constraint $\Eq~[\aA]$ to $\Eq~\aA$, yielding
the signature $\blueforall \aA.~\Eq~\aA \Rightarrow \aA \to \Bool$.


Second, evidence construction is not performed directly, by means of creating a
dictionary. Instead, a dictionary substitution $\evSubst$ is created, which
maps wanted dictionary variables to dictionaries. This strategy is analogous
to the unification algorithm, which solves type equalities by creating a type
substitution for instantiating the yet-unknown types.

Finally, algorithmic constraint entailment does not take the complete program
theory as the specification does, but an axiom set. We make this design choice
due to superclass constraint schemes: during simplification we do not want to
replace a wanted constraint $(\Eq~\aA)$ with a more complex $(\Ord~\aA)$. We
elaborate on the transition from the program theory $\programTheory$ to an
equally expressive axiom set $\axiomSet$---which does not contain superclass
constraint schemes---next.

\subsubsection{Transitive Closure of the Superclass Relation}
\label{subsubsec:transitive-closure}

Superclass axioms often overlap with instance axioms. Consider for example the
following two axioms, the first obtained by the $\Eq$ instance for lists and
the second obtained by the $\Ord$ class declaration:
\[\small
\begin{array}{l@{\qquad\qquad}l}
  \blueforall \aA.~\Eq~\aA \Rightarrow \Eq~[\aA] & \text{(a)} \\
  \blueforall \bB.~\Ord~\bB \Rightarrow \Eq~\bB  & \text{(b)} \\
\end{array}
\]
This is a problem for type inference, since the constraint solving algorithm
would have to make a choice when faced for example with constraint $\Eq~[\cC]$.
Both (a) and (b) match but to completely entail constraint $\Eq~[\cC]$ we would
require $\Eq~\cC$ if we were to choose the former and $\Ord~[\cC]$ if we were
to choose the latter.
In order to avoid this source of non-determinism, several implementations of
type classes (and notably GHC) treat superclass constraints differently.

In essence, we can pre-compute the transitive closure of the superclass
relation on a set of given constraints and omit superclass axioms altogether.
This procedure should also be reflected in the elaborated terms. To
this end, we introduce dictionary contexts $\matchCtx$: 
\[\small
\begin{array}{@{\hspace{0mm}}c@{\hspace{0mm}}}
\begin{tabular*}{0.9\columnwidth}{@{\hspace{0mm}}l@{\hspace{1mm}}c@{\hspace{1mm}}l@{\extracolsep{\fill}}r@{\hspace{0mm}}}
  \syntaxLine{\matchCtx}
             {::=}
             {\square \mid \keylet~d : \fcType = t~\keyin~\matchCtx}
             {dictionary context}
\end{tabular*}
\end{array}
\]
During entailment we can replace the program theory $\programTheory$ with an
axiom set $\axiomSet$ which does not contain any superclass axioms and a
dictionary context $\matchCtx$. This procedure we denote as
$\scClosure{\as}{\programTheory} = (\axiomSet,\matchCtx)$:
\[\small
\begin{array}{l}
  \scClosure{\as}{\langle\superAxiomSet,\instanceAxiomSet,\annConstraintSet_L\rangle} = ((\instanceAxiomSet, \annConstraintSet_L', \annConstraintSet_L), \matchCtx) \\
  ~~\text{where}~ (\annConstraintSet_L', \matchCtx) = \mpClosure{\as}{\superAxiomSet}{\annConstraintSet_L}
\end{array}
\]
Function $\mathit{closure}$ computes the transitive closure of the following
function:
\[\small
\begin{array}{@{\hspace{0mm}}l@{\hspace{1mm}}c@{\hspace{1mm}}l@{\hspace{0mm}}}
  \multicolumn{3}{@{\hspace{0mm}}l@{\hspace{0mm}}}{
  \mpOneStep{\as}{\axiomSet}{d : \TC~\monotype_1} = (\mathit{bimap}~\mathit{mconcat}~\mathit{mconcat}\cdot\mathit{unzip})} \\
  \quad & \{   &(\{d_2 : \tySubst(\constraint)\}, \matchCtx) \\
        & \mid & (d_1 : \blueforall \bs.~\TC~\monotype_1 \Rightarrow
                 \constraint) \in \axiomSet \\
        & ,    & \tcUnify{\as}{\monotype_1 \sim \monotype_2} = \tySubst \\
        & ,    & \matchCtx = \keylet~d_2 : \tcElabCt{\tySubst(\constraint)} =
                 d_1~\tySubst(\bs)~d~\keyin~\square \\
        & \} \\
\end{array}
\]
Function $\mpOneStep{\as}{\axiomSet}{\annConstraint} = (\annConstraintSet,
\matchCtx)$ tries to match the left-hand side of every available constraint
scheme in $\axiomSet$ with the given constraint. If matching is successful,
modus ponens is used to derive the right-hand side. This procedure is also
reflected in the dictionary context $\matchCtx$, which captures a scope where
the derived dictionaries are available.
For example, if 
\[\small
\begin{array}{l@{\hspace{1mm}}c@{\hspace{1mm}}l}
  \as            & = & m \\
  \superAxiomSet & = & \{~d_1 : \blueforall n.~\Monad~n \Rightarrow \Applicative~n \\
                 &   & ,~d_2 : \blueforall k.~\Applicative~k \Rightarrow \Functor~k~\} \\
  \annConstraint & = & d_3 : \Monad~m \\
\end{array}
\]
then $\mpClosure{\as}{\superAxiomSet}{\annConstraint}$ results in the following:
\[\small
\begin{array}{l@{\hspace{1mm}}c@{\hspace{1mm}}l}
  \axiomSet  & = & \{~d_4 : \Applicative~m, d_5 : \Functor~m~\} \\
  \matchCtx  & = & \keylet~d_4 : \tyCon_\Applicative~m = d_1~m~d_3~\keyin \\
             &   & \keylet~d_5 : \tyCon_\Functor~m = d_2~m~d_4~\keyin~\square \\
\end{array}
\]
In plain type inference, superclasses are never used; the above procedure is
required in type {\em checking}. This is the case for method implementations,
explicitly-annotated terms, and the entailment of superclass constraints in
class instances.
This is better illustrated in the elaboration of declarations, which we discuss
next.

\subsubsection{Declaration Elaboration}

We now turn to type inference and elaboration for top-level declarations. Since
type inference with elaboration for class declarations is identical to its
specification, we only discuss the judgments for class instances and top-level
bindings.

\paragraph{Instance Inference with Elaboration}

Typing inference for instance declarations takes the form
$\tcElabInsNoShade{\programTheory}{\tyEnv}{\insDecl}{\programTheory'}{\decl}$
and is given by the following rule:
\begin{mathpar}
\small
\inferrule*[right=]
           { 
             \keyclass~\blueforall a.~\supconstraints^n \Rightarrow \TC~a~\keywhere~\{~f :: \polytype~\} \\
             \tyEnv_I = \tyEnv, \bs \\
             \programTheory_I = \programTheory\lComma \overline{\highlight{d :} \constraint}^m \\
             \tcTcCt{\tyEnv_I}{\overline{\constraint}^m}{\fcTypes^m} \\
             \tcTcTy{\tyEnv_I}{\monotype}{\fcType} \\
             \constraintScheme = \blueforall \bs.~\overline{\constraint}^m \Rightarrow \TC~\monotype \\
             \scClosure{\bs}{\programTheory_I} = (\axiomSet,\highlight{\matchCtx}) \\
             \tcSimplifyAllCt{\bs}{\axiomSet}{\overline{\highlight{d :} \constraint}^n}{\bullet}{\eta} \\
             %
             \tcSubsumption{\bs}{\programTheory_I\iComma \highlight{d :} \constraintScheme}{\tyEnv_I}{e}{[\monotype/a]\polytype}{t} \\
             \highlight{\decl = \keylet~d : \blueforall \bs.\fcTypes_i^m \to \tyCon_\TC~\fcType = \Lambda \bs.~\lambda (\overline{d : \fcType}^m).~\dataCon_\TC~\fcType~\applyExprCtx{\matchCtx}{\evSubst(\supdicts^n)}~t}
           } 
           { \tcElabIns{\programTheory}
                       {\tyEnv}
                       {\keyinstance~\blueforall \bs.\overline{\constraint}^m \Rightarrow \TC~\monotype~\keywhere~\{f = e\}}
                       {[\highlight{d :} \constraintScheme]}
                       {\decl}
           }
\end{mathpar}
For the most part it is identical to the corresponding rule in
Figure~\ref{fig:basic-system-specification}. The most notable differences are
concentrated around superclass entailment and type checking of the method
implementation.

For the entailment of the superclass constraints we pre-compute the transitive
closure of the superclass relation, and then
\begin{inparaenum}[(a)]
\item
  we generate fresh dictionary variables $\supdicts^n$, to capture the yet-unknown
  superclass dictionaries, and
\item
  we exhaustively simplify the superclass constraints (requiring no residual
  constraints), obtaining an evidence substitution $\evSubst$. $\evSubst$ maps
  dictionary variables $\supdicts^n$ to generated dictionaries; the complete witness
  for the $i$-th superclass dictionary takes the form
  $\applyExprCtx{\matchCtx}{\evSubst(\tySubst(\supdict_i))}$.
\end{inparaenum}

Method implementations have their type imposed by their signature in the
class declaration. Hence, we need to {\em check} rather than {\em infer} their
type.
This operation is expressed succinctly by relation
$\tcSubsumptionNoShade{\as}{\programTheory}{\tyEnv}{e}{\polytype}{t}$: 
\begin{mathpar}
\small
\inferrule*[right=]
           { \tcElabTm{\tyEnv}{e}{\monotype_1}{t}{\annConstraintSet}{\equalities} \\
             \tySubst = \tcUnify{\as, \bs}{\equalities, \monotype_1 \sim \monotype_2} \\
             \scClosure{\as}{(\programTheory\lComma \overline{\highlight{d :} \constraint}^n)} = (\axiomSet,\highlight{\matchCtx}) \\
             \tcSimplifyAllCt{\as, \bs}
                             {\axiomSet}
                             {\tySubst(\annConstraintSet)}{\bullet}{\evSubst}
           } 
           { \tcSubsumption{\as}
                           {\programTheory}
                           {\tyEnv}
                           {e}
                           {(\blueforall \bs.~\constraints^n \Rightarrow \monotype_2)}
                           {\Lambda \bs.~\lambda \overline{(d : \tcElabCt{\constraint})}^n.~\applyExprCtx{\matchCtx}{\evSubst(\tySubst(t))}}
           } 
\end{mathpar}
Essentially, it ensures that the inferred type for $e$ {\em subsumes} the
expected type $\polytype$. A type $\polytype_1$ is said to subsume type
$\polytype_2$ if any expression that can be assigned type $\polytype_1$ can
also be assigned type $\polytype_2$.
%
The above rule performs type inference and type subsumption checking
simultaneously:
First, it infers a monotype $\monotype_1$ for expression $e$, as well as wanted
constraints $\annConstraintSet$ and type equalities $\equalities$. Type
equalities $\equalities$ should have a unifier and the inferred type
$\monotype_1$ should also be unifiable with the expected type $\monotype_2$.
Finally, the given constraints $\overline{\constraint}^n$ should completely
entail the wanted constraints $\annConstraintSet$. For constraint entailment,
we (again) pre-populate the given constraints with the transitive closure of
the superclass axioms.

\paragraph{Value Binding Inference with Elaboration}

Finally, type inference for top-level bindings is given by the judgment
$\tcElabValNoShade{\programTheory}{\tyEnv}{\valDecl}{\tyEnv'}{\decl}$.
%
The first rule deals with annotation-free bindings:

\begin{mathpar}
\small
\inferrule*[right=]
           { \tcElabTm{\tyEnv, x : b}{e}{\monotype}{t}{\annConstraintSet}{\equalities} \\
             \tcUnify{\bullet}{\equalities, b \sim \monotype} = \tySubst \\
             \as = \fv{\tySubst(\annConstraintSet)} \cup \fv{\tySubst(\monotype)} \\
             \tcSimplifyAllCt{\as}{\instanceAxiomSet,\annConstraintSet_L}{\tySubst(\annConstraintSet)}{(\overline{\highlight{d :} \constraint}^n)}{\evSubst} \\
             \highlight{\fcType = \tcElabTy{\blueforall \as.~\overline{\constraint}^n \Rightarrow \tySubst(\monotype)}} \\
             \metaforall {\constraint_i \in \constraints^n}
               \highlight{\fcType_i = \tcElabCt{\constraint_i}} \\
             \highlight{\decl = \keylet~x : \fcType = \Lambda \as.~\lambda (\overline{d : \fcType}^n).~\applyExprCtx{\matchCtx}{\evSubst(\tySubst(t))}}
           } 
           { \tcElabVal{\langle\superAxiomSet,\instanceAxiomSet,\annConstraintSet_L\rangle}
                       {\tyEnv}
                       {x = e}
                       {[x : \blueforall \as.~\overline{\constraint}^n \Rightarrow \tySubst(\monotype)]}
                       {\decl}
           } 
\end{mathpar}
The rule performs constraint generation, simplification, and generalization of
an annotation-free top-level binding. Though straightforward, it is worth
noticing that superclass axioms $\superAxiomSet$ are ignored, since there are
no local (given) constraints.

The second rule deals with explicitly annotated bindings:
\begin{mathpar}
\small
\inferrule*[right=]
           { \tcSubsumption{\bullet}{\programTheory}{\tyEnv, x : \polytype}{e}{\polytype}{t} }
           { \tcElabVal{\programTheory}{\tyEnv}{\keylet~x : \polytype = e}{[x : \polytype]}{\keylet~x : \tcElabTy{\polytype} = t} }
\end{mathpar}
Essentially, type inference for annotated terms directly corresponds to an
inference-and-subsumption-check, as given by judgment
$\tcSubsumptionNoShade{\as}{\programTheory}{\tyEnv}{e}{\polytype}{t}$ above.

\section{Bidirectional Instances, Formally}\label{sec:the-basic-system}

In this section we present the changes needed for extending the basic system of
Section~\ref{sec:basic-with-superclasses} with support for bidirectional
instances.

\subsection{Syntax Extensions}\label{sec:classes-bidirectional-syntax}

First, in order to use the inverted axioms selectively and avoid the
termination issue we mentioned in
Section~\ref{sec:classes-bidirectional-challenges:termination}, we extend the
syntax of program theory $\programTheory$ with an additional component, the
inverted instance axioms $\invertedAxiomSet$:\footnote{Subscript $B$ stands for
``Bidirectional''.}
\[\small
\begin{array}{@{\hspace{0mm}}c@{\hspace{0mm}}}
\begin{tabular*}{0.9\columnwidth}{@{\hspace{0mm}}l@{\hspace{1mm}}c@{\hspace{1mm}}l@{\extracolsep{\fill}}r@{\hspace{0mm}}}
  \syntaxLine{\programTheory}
             {::=}
             {\langle \highlight{\invertedAxiomSet,} \superAxiomSet, \instanceAxiomSet, \localTheory \rangle}
             {program theory}
\end{tabular*}
\end{array}
\]
As we illustrate below---similarly to superclass axioms
$\superAxiomSet$---inverted instance axioms $\invertedAxiomSet$ are used for
type checking but not for type inference. The rest of the syntax is identical
to the syntax of the basic system we presented in
Figure~\ref{subfig:class-background:tc-syntax}.

\subsection{Specification Extensions}\label{sec:classes-bidirectional-specification}

The specification of typing and elaboration is for the most part identical to
that of Section~\ref{sec:the-basic-system:specifications}. The changes
bidirectional instances introduce are concentrated in class and instance
declaration typing, which we now discuss.

\subsubsection{Class Declarations}

The specification of class typing with elaboration -- as well as for the basic
system -- is given by judgment
$\bisTcClsDeclNoShade{\tyEnv}{\clsDecl}{\programTheory}{\tyEnv'}{\overline{\decl}}$
(Figure~\ref{fig:basic-system-specification}). Since
for the most part the rule is identical to the basic system, we only highlight
the differences. For a class declaration of the form
\[\small
\keyclass~\blueforall a.~\constraints^n \Rightarrow \TC~a~\keywhere~\{~f :: \polytype~\}
\]
we have the following:

\paragraph{1. Instance Context}

Firstly, in addition to the superclass and method projections, the class
declaration gives rise to a \SystemFC open type family declaration:
\[\small
\keytype~\tyFam_\TC~a
\]
Function $\tyFam_\TC~a$ captures the functional dependency between the instance
context and the class parameter. Hence, function $\tyFam_\TC$ is populated by
$\TC$ instances, each mapping its class parameter to the corresponding instance
context. 


\paragraph{2. Dictionary Representation}

Secondly, we extend the dictionary declaration, so that it can store the
instance context of type $\tyFam_\TC~a$:\footnote{
  The order of the dictionary arguments is irrelevant, and the choice made here
  is arbitrary.
}
\[\small
\keydata~\tyCon_\TC~a = \dataCon_\TC~\highlight{(\tyFam_\TC~a)}~\fcTypes^n~\fcType
\]

\paragraph{3. Projection Functions}

Finally, since the data constructor $\dataCon_\TC$ now stores an additional
field, we ``shift'' the superclass and method projections accordingly:
\[\small
\begin{array}{l}
  \keylet~d_i : \blueforall a.~\tyCon_\TC~a \to \fcType_i = \Lambda a.~\lambda (d : \tyCon_\TC~a).~\genProject{i+1}{\TC}{d} \\ 
  \keylet~f : \blueforall a.~\tyCon_\TC~a \to \fcType   = \Lambda a.~\lambda (d : \tyCon_\TC~a).~\genProject{n+2}{\TC}{d}   \\
\end{array}
\]

\subsubsection{Instance Declarations}


Typing for instance declarations also preserves the signature we gave in
Figure~\ref{fig:basic-system-specification}. For a class instance of the form
\[\small
\keyinstance~\blueforall \bs.~\constraints^m \Rightarrow \TC~\monotype~\keywhere~\{~f = e~\}
\]
bidirectional instances introduce the following extensions:

\paragraph{1. Instance Context Axiom}

Firstly, an additional clause is generated for function $\tyFam_\TC$, capturing
the dependency between the instance parameter $\monotype$ and the instance
context:
\[\small
\keyaxiom~\axiom{\TC}{\monotype}~\bs : \tyFam_\TC~\fcType \sim (\fcType_1, \ldots, \fcType_m)
\]
where $\fcType_i$ is the dictionary type representation of $\constraint_i$ in
the instance context and $\fcType$ is the elaboration of parameter $\monotype$.

\paragraph{2. Inverted Instance Axioms}

Secondly, the program theory extension introduced by the instance now includes
the inverted instance axioms, which take the form:
\[\small
\constraintScheme_i = \blueforall \bs.~\TC~\monotype \Rightarrow \constraint_i \qquad i \in [1\dots m]
\]
Of course, such implications need to be reflected in term-level functions in
the generated \SystemFC code. Hence, for every implication
$\constraintScheme_i$, we generate a projection function $d_i$, given by the
following definition:
\[\small
\begin{array}{@{\hspace{0mm}}l@{\hspace{1mm}}c@{\hspace{1mm}}l@{\hspace{0mm}}}
  \keylet~d_i & : & \blueforall \bs.~\tyCon_\TC~\fcType \to \fcType_i \\
              & = & \Lambda \bs.~\lambda (d : \tyCon_\TC~\fcType).~\keycase~d~\keyof \\
              &   & \quad\dataCon_\TC~\mathit{ctx}~\ds^n~x \to \keycase~\fcCast{\mathit{ctx}}{(\axiom{\TC}{\monotype}~\bs)}~\keyof~(d_1,...,d_m) \to d_i \\
\end{array}
\]
The outer pattern matching exposes the instance context $\mathit{ctx}$, of type
$\tyFam_\TC~\fcType$, which we explicitly cast to a tuple of all instance
context dictionaries $(d_1, \ldots, d_m)$. Then, the inner pattern matching
extracts and returns the corresponding instance context dictionary $d_i$.

\paragraph{3. Storing the Instance Context}

Finally, the implementation of the instance dictionary (transformer) needs to
store the instance context dictionaries within the dictionary for
$\TC~\monotype$. Thus, the instance dictionary (transformer) now takes the
form:
\[\small
\begin{array}{l@{\hspace{1mm}}c@{\hspace{1mm}}l}
  \keylet~d & : & \blueforall \bs.~\fcTypes_i^m \to \tyCon_\TC~\fcType \\
            & = & \Lambda \bs.~\lambda (\overline{d : \fcType}^m).~\dataCon_\TC~\fcType~\highlight{(\fcCast{(d_1, \ldots, d_m)}{\fsym{(\axiom{\TC}{\monotype}~\bs)}})}~\ts^n~t] \\
\end{array}
\]
For the constructed
dictionary to be well-typed, the tuple $(d_1, \ldots, d_m)$ containing all
instance context dictionaries needs to be explicitly cast to have type
$\tyFam_\TC~\fcType$, as the type of $\dataCon_\TC$ requires. This is exactly
what $\coercion$ proves:
$(\fcType_1, \ldots, \fcType_m) \sim \tyFam_\TC~\fcType$.
%

\subsection{Algorithm Extensions}\label{sec:classes-bidirectional-algorithm}

Type inference is again for the most part identical to that of the basic system
(Section~\ref{sec:the-basic-system:algorithms}). The changes bidirectional
instances introduce are concentrated in declarations. Type inference for classes
is identical to its specification so we only discuss the
differences in class instances and value bindings.

\paragraph{Instance Declarations}

Type inference for instance declarations behaves similarly to its
specification. The main difference lies in the type inference and subsumption
checking for methods.

In addition to the superclass closure we also compute the transitive closure of
the inverted axioms. Thus, we replace function $\scClosure{\as}{\programTheory}
= (\axiomSet,\matchCtx)$ of Section~\ref{subsubsec:transitive-closure} with
$\invScClosure{\as}{\programTheory} = (\axiomSet,\matchCtx)$:
\[\small
\begin{array}{l}
  \invScClosure{\as}{\langle\invertedAxiomSet,\superAxiomSet,\instanceAxiomSet,\annConstraintSet_L\rangle} = ((\annConstraintSet_L', \instanceAxiomSet, \annConstraintSet_L), \matchCtx) \\
  ~~\text{where}~ (\annConstraintSet_L', \matchCtx) = \mpClosure{\as}{(\invertedAxiomSet,\superAxiomSet)}{\annConstraintSet_L}
\end{array}
\]

\paragraph{Value Bindings}

Type inference for value bindings is also mildly affected by bidirectional
instances. Bindings without a type annotation ignore the
inverted axioms, alongside the superclass axioms; users that enable
bidirectional instances can expect type inference to behave as usual.

A top-level value binding with an explicit type annotation behaves differently,
similarly to method typing. Just as we do with superclass constraints, we also
compute the transitive closure of the inverted axioms, making more derivations
possible. This extension---along with all the changes we described in this
section---manifests itself in the elaboration of $\cmp_2$
(Section~\ref{sec:classes-bidirectional-formal-extensions}), as dictionaries
$\dD_1'$ and $\dD_2'$.

\section{Meta-theory}\label{sec:classes-bidirectional-metatheory}

Since the formalization of bidirectional instances conservatively extends that
of Section~\ref{sec:the-basic-system}, we focus on the most interesting
meta-theoretical properties of our extension: termination of type inference and
the principal type property.

\subsection{Termination}
\label{sec:classes-bidirectional-metatheory:termination}

\paragraph{Termination Conditions}

First, for decidable type inference it is required that type inference
terminates on all inputs.
The following {\em Termination Conditions} are sufficient to ensure termination
of type inference:
\begin{enumerate}[(a)]
\item
  The superclass relation forms a {\em Directed Acyclic Graph}.
\item
  In each class instance ($\keyinstance~\blueforall \bs.~\constraintSet \Rightarrow
  \TC~\monotype$):
  \begin{itemize}
  \item
    no variable has more occurrences in a type class constraint in the instance
    context $\constraintSet$ than the head $(\TC~\monotype)$,
  \item
    each class constraint in the instance context $\constraintSet$ has fewer
    constructors and variables (taken together, counting repetitions) than the
    head $(\TC~\monotype)$.
  \end{itemize}
\end{enumerate}
The first restriction ensures that the computation of the transitive closure of
the superclass relation 
is terminating~\citep[Sec.~4.3.1]{haskell98}.
The second restriction~\citep[Def.~11]{fundeps-chr} ensures that instance
contexts are {\em decreasing}, so that class resolution is also terminating.
To illustrate why type inference in the presence of bidirectional instances
terminates, we first distinguish between type inference and type checking.

\paragraph{Termination of Type Inference}

In cases where a type is inferred, the algorithm is identical to that of the
basic system; the feature manifests itself when there are 
type signatures. Hence, in these cases decreasing instance contexts are sufficient
to ensure termination.

\paragraph{Termination of Type Checking}

In cases where we need to check an expression against a type, the inverted
axioms also come into play, as well as the superclass axioms.
Since we compute the closure of the superclass relation and the inverted axioms
(by means of function $\invScName$), we need to ensure that both superclass and
inverted axioms cannot be applied indefinitely.
For the former, Condition~(a) is sufficient: any uninterrupted sequence of
superclass axiom applications is bounded by the height of the superclass graph.
For the latter, decreasing contexts are also sufficient. To illustrate why,
consider the following inverted axiom:
\[\small
\blueforall a.~\blueforall b.~\Eq~(a, b) \Rightarrow \Eq~a
\]
During completion, $\invScName$ applies the axiom to constraints of the form
$\Eq~(\monotype_1, \monotype_2)$, ending up with an additional axiom of a
smaller size: $\Eq~\monotype_1$.
In short, uninterrupted sequences of inverted axioms are bounded by the size of
the types in instance heads.
%
In short, any step $\invScName$
takes either reduces the size of a constraint, or takes a step in the
superclass graph, both of which are bounded.

\subsection{Principality of Types}
\label{sec:classes-bidirectional-metatheory:principality}

Our specification
(Sections~\ref{sec:the-basic-system:specifications}
and~\ref{sec:classes-bidirectional-specification}) possesses the principal type
property: the definition of a principal type does not specify one type, but
rather the properties of it. That is, the following types of $\cmp$
(Section~\ref{sec:classes-bidirectional-challenges}) are both equally general: %
\[\small
\begin{array}{l@{\hspace{1mm}}c@{\hspace{1mm}}l@{\hspace{1mm}}c@{\hspace{1mm}}l}
  \cmp & :: & \blueforall \aA.~ \Eq~\aA   & \Rightarrow & \aA \to \aA \to \Bool \\
  \cmp & :: & \blueforall \aA.~ \Eq~[\aA] & \Rightarrow & \aA \to \aA \to \Bool \\
\end{array}
\]
Hence, the main concern is whether the type inference algorithm of
Sections~\ref{sec:the-basic-system:algorithms}
and~\ref{sec:classes-bidirectional-algorithm} infers one of the principal
types. The answer is {\em yes}.
Since plain type inference does not exploit the inverted axioms, the algorithm
infers backwards-compatible principal types. Backwards-chaining simplifies
constraints such as $\Eq~[\aA]$ to $\Eq~\aA$ but not the other way around.
Thus, the algorithm would never infer type 
\[\small
\blueforall \aA.~\blueforall \bB.~\Eq~(\aA, \bB) \Rightarrow \ldots
\]
but would infer the isomorphic (and also principal) type
\[\small
\blueforall \aA.~\blueforall \bB.~(\Eq~\aA, \Eq~\bB) \Rightarrow \ldots
\]
Expressions with explicit type annotations have only one principal type: the
one specified by their signature. In these cases the algorithm will use the
inverted axioms to entail the wanted constraints $(\Eq~\aA, \Eq~\bB)$ using the
given $\Eq~(\aA, \bB)$, thus constructing again the principal type.

That is, in the absence of type annotations the principal type is the principal
Haskell98 type, and in the presence of type annotations the type annotation
dictates what the principal type is.
In either case, our algorithm reconstructs the principal type, therefore
addressing the challenge of
Section~\ref{sec:classes-bidirectional-challenges:principal-types}.

\subsection{Other Properties}
\label{sec:classes-bidirectional-metatheory:other}

\paragraph{Preservation of Typing Under Elaboration}

We are confident that the specification of elaboration we gave in
Sections~\ref{sec:the-basic-system:specifications}
and~\ref{sec:classes-bidirectional-specification} is type-preserving. The
formal proof of this statement we leave for future work.


\paragraph{Soundness of Generated Code}

It is known that
overlapping instances make the semantics of type classes incoherent but they do
not introduce unsoundness. In the presence of bidirectional instances, this is
no longer true:
%
\[\small
\begin{array}{@{\hspace{0mm}}l@{\hspace{2mm}}c@{\hspace{2mm}}l@{\hspace{0mm}}}
  \keyinstance~\Eq~\aA \Rightarrow \mathit{Eq}~[\aA] & \rightsquigarrow & \keyaxiom~g_1~\aA : \tyFam_{\Eq}~[\aA] \sim \tyCon_{\Eq}~\aA \\
  \keyinstance~\Eq~[\bB]                             & \rightsquigarrow & \keyaxiom~g_2~\bB : \tyFam_{\Eq}~[\bB] \sim \Unit            \\
\end{array}
\]
Axioms $g_1$ and $g_2$ violate the \SystemFC~{\em compatibility
condition}~\citep[Defn.~10]{closedtf}, which means that our elaboration would
give rise to unsound \SystemFC code. Indeed,
$\fcomp{(\fsym{(g_1~\Int)})}{(g_2~\Int)}$ is a proof of $\tyCon_{\Eq}~\Int \sim
\Unit$.
We revisit this issue in Section~\ref{sec:classes-bidirectional:related}.

\paragraph{Coherence}

In the absence of overlapping instances and ambiguous types, we conjecture that our elaboration is
{\em coherent}. Given the similarity between the handling of superclass
constraints and bidirectional instances, we are confident that the recent
advances of~\citet{CoherenceProof} could be easily extended to accommodate
bidirectional instances.

\paragraph{Algorithm Soundness and Completeness}

Finally, we conjecture that the algorithm of Sections~\ref{sec:the-basic-system:algorithms} and~\ref{sec:classes-bidirectional-algorithm}
is sound and complete with respect to its
specification.

\section{Related Work and Discussion}\label{sec:classes-bidirectional:related}

\paragraph{Class Elaboration}

Maybe the most relevant line of work is the
specification
of typing and elaboration (into \SystemF) of type classes with superclasses, given by~\citet{ClassElaboration}.
%
Yet, the work of~\citeauthor{ClassElaboration} does not cover an algorithm for type
inference and elaboration; we do so here
(Section~\ref{sec:the-basic-system:algorithms}).

\paragraph{Constrained Type Families}

\citet{constrained-type-families} recently provided compelling
arguments for the replacement of open type families with the so-called {\em
Constrained Type Families}. Constrained type families, similarly to associated
type families, use the generic notion of qualified types~\citep{qualtypes} to
capture the domain of a type family within a predicate, thus simplifying the
meta-theory of type families and their extensions.

Within this setting, the bidirectionality of the axioms is essential. Indeed,
\citeauthor{constrained-type-families} use a variation of the the $\append$
example (Section~\ref{sec:classes-bidirectional-motivation}) to motivate the extension of \SystemFC with the
$\texttt{assume}$ construct, which axiomatically provides the bidirectionality
needed for $\append$ to type check. Unfortunately, $\texttt{assume}$ is not a
panacea: axiomatically assuming the satisfiability of constraints does not
scale to class methods.\footnote{This is also the case for other attempts to
tackle the same problem using other type-level features.  See for example
\url{http://okmij.org/ftp/Haskell/number-parameterized-types.html\#binary-arithm}.}

\paragraph{Overlapping Instances}

As we mentioned in Section~\ref{sec:classes-bidirectional-metatheory:other},
bidirectional instances can lead to unsound \SystemFC code in the presence of
the (in)famous \texttt{OverlappingInstances} GHC extension.
Though this extension is considered harmful---and has thus been deprecated
since GHC 7.10 in favour of more fine-grained per-instance
pragmas---it is still
used, making it important to study its interaction with our feature. 

Depending on the level of overlap allowed, we can selectively make instances
bidirectional: the system is sound if overlap and bidirectionality are
aligned. Indeed, instances determine the generated axioms so our strategy is
simple: {\em any instance that overlaps with other instances should not give
rise to any inverted axioms}.

In terms of the overlapping $\Eq$ instances of the previous section, this means
that we would give rise to
\[\small
\begin{array}{@{\hspace{0mm}}l@{\hspace{2mm}}c@{\hspace{2mm}}l@{\hspace{0mm}}}
  \keyinstance~\Eq~\aA \Rightarrow \mathit{Eq}~[\aA] & \rightsquigarrow & \keyaxiom~g_1~\aA : \tyFam_{\Eq}~[\aA] \sim \Unit \\
  \keyinstance~\Eq~[\bB]                             & \rightsquigarrow & \keyaxiom~g_2~\bB : \tyFam_{\Eq}~[\bB] \sim \Unit \\
\end{array}
\]
thus ensuring safety of the generated code.



\paragraph{Instance Chains}

Though our design generates ``open'' equality axioms (to agree with the open
nature of type classes), one might also consider bidirectionality in the
presence of {\em``instance chains''}~\citep{InstanceChains}. Instance chains
allow for ordered overlap among instances, which we believe can be
combined with our interpretation.
Instead of a collection of open axioms, an instance chain can give rise to a
``closed'' equality axiom (like the ones generated by {\em closed type
families}~\citep{closedtf}), to preserve soundness of the generated code
without sacrificing expressive power.

\paragraph{Inversion Principles in Proof Assistants}

There is also a large body of work concerned with {\em inversion principles},
with significant applications in the area of proof assistants (see for example
tactic
\href{https://coq.inria.fr/refman/proof-engine/tactics.html\#coq:tacn.inversion}{\texttt{inversion}}).
Though inversion principles seem like a more natural approach for addressing
the problem we target here, the open nature of type classes disallows a direct
application to Haskell. Nevertheless, we would like to explore this alternative
approach in the future.

%
%
%
%
%
%
%
%
%

\paragraph{Denotational Semantics for Type Classes}

\citet{SimpleSemanticsForHaskellOverloading} gives an---inherently
bidirectional---denotational semantics for type classes, rather than through a
dictionary-passing translation. Within this work, polymorphic instances are
interpreted extensionally, as the set of their ground instantiations.
Unfortunately, it has not been studied yet how this semantics
relates to the traditional dictionary-based semantics that we target here.

\paragraph{Quantified Class Constraints}

An interesting avenue for future work is studying the interaction between {\em
Quantified Class Constraints}~\citep{quantcs} and Bidirectional Instances. The
two key challenges are (a) elaboration and (b) type inference.

Combining the elaboration strategies of the features is a
straightforward task. For example, the instance
\[
\keyinstance~(\blueforall \aA.~\Monoid~(\fF~\aA)) \Rightarrow \Alternative~\fF
\]
generates the following \SystemFC axiom:\footnote{Notice though that this
encoding needs more \SystemFC power than GHC currently uses; it is impossible
to encode Bidirectional Instances combined with Quantified Class Constraints
using the current GHC version.}
\[
  \keyaxiom~g~\fF : \tyFam_\Alternative~\fF \sim \blueforall \aA.~\tyCon_\Monoid~(\fF~\aA)
\]

The second aspect, type inference, is more interesting. The main challenge lies
in the significantly different constraint entailment strategies: Quantified
Class Constraints use backtracking to ensure completeness, but Bidirectional
Instances can lead to non-termination in the presence of backtracking (see
Section~\ref{sec:classes-bidirectional-challenges:termination}).
We believe that a restricted combination of the two features is possible,\footnote{GHC also
supports a limited version of Quantified Constraints (see commit
\href{https://gitlab.haskell.org/ghc/ghc/commit/7df589608abb178efd6499ee705ba4eebd0cf0d1}{7df589608abb178efd6499ee705ba4eebd0cf0d1}),
without backtracking.} and plan to investigate their interaction in the future.


\section{Conclusion}\label{sec:conclusion}

We have presented a conservative extension of type classes, which allows class
instances to be interpreted bidirectionally, thus significantly improving the
interaction of GADTs with type classes, by allowing proper structural induction
over GADTs, even in the presence of qualified types.

\appendix

\section{Additional Judgments}
\label{appendix:remaining-specification}

\subsection{Specification of Typing and Elaboration}

\begin{figure}
\small
\begin{flushleft}
\namedRuleform{\tcTcTy{\tyEnv}{\polytype}{\fcType}}
              {Type Well-formedness with Elaboration}
\end{flushleft}
\begin{mathpar}
\inferrule*[right=]
           {a \in \tyEnv}
           {\tcTcTy{\tyEnv}{a}{a}}

\inferrule*[right=]
           {\tcTcTy{\tyEnv}{\monotype_1}{\fcType_1} \\
            \tcTcTy{\tyEnv}{\monotype_2}{\fcType_2}
           } 
           {\tcTcTy{\tyEnv}{\monotype_1 \to \monotype_2}{\fcType_1 \to \fcType_2}}

\inferrule*[right=]
           {\tcTcTy{\tyEnv, a}{\polytype}{\fcType}}
           {\tcTcTy{\tyEnv}{\blueforall a.~\polytype}{\blueforall a.~\fcType}}

\inferrule*[right=]
           {\tcTcCt{\tyEnv}{\constraint}{\fcType_1} \\
            \tcTcTy{\tyEnv}{\qualtype}{\fcType_2}
           } 
           {\tcTcTy{\tyEnv}{\constraint \Rightarrow \qualtype}{\fcType_1 \to \fcType_2} }
\end{mathpar}

\begin{flushleft}
\namedRuleform{\tcTcCt{\tyEnv}{\constraint}{\fcType}}
              {Constraint Well-formedness}
\end{flushleft}
\begin{mathpar}
\inferrule*[right=ClsCt]
           { \TC~\text{defined} \\
             \tcTcTy{\tyEnv}{\monotype}{\fcType}
           } 
           {\tcTcCt{\tyEnv}{\TC~\monotype}{\tyCon_\TC~\fcType}}
\end{mathpar}

\begin{flushleft}
\namedRuleform{\tcTcTm{\programTheory}{\tyEnv}{e}{\polytype}{t}}
              {Term Typing with Elaboration}
\end{flushleft}
\begin{mathpar}
\inferrule*[right=Var]
           {(x : \polytype) \in \tyEnv}
           {\tcTcTm{\programTheory}{\tyEnv}{x}{\polytype}{x}}

\inferrule*[right=$(\blueforall\hspace{0mm}I)$]
           {\tcTcTm{\programTheory}{\tyEnv, a}{e}{\polytype}{t}}
           {\tcTcTm{\programTheory}{\tyEnv}{e}{\blueforall a.~\polytype}{\blueforall a.~t}}

\inferrule*[right=$(\blueforall\hspace{0mm}E)$]
           {\tcTcTm{\programTheory}{\tyEnv}{e}{\blueforall a.~\polytype}{t} \\
            \tcTcTy{\tyEnv}{\monotype}{\fcType}
           } 
           {\tcTcTm{\programTheory}{\tyEnv}{e}{[\monotype/a]\polytype}{t~\fcType}}

\inferrule*[right=$(\rightarrow\hspace{-1mm}I)$]
           {\tcTcTy{\tyEnv}{\monotype_1}{\fcType_1} \\
            \tcTcTm{\programTheory}{\tyEnv, x : \monotype_1}{e}{\monotype_2}{t}
           } 
           {\tcTcTm{\programTheory}{\tyEnv}{\lambda x.~e}{\monotype_1 \to \monotype_2}{\lambda (x : \fcType_1).~t_2}}

\inferrule*[right=$(\rightarrow\hspace{-1mm}E)$]
           {\tcTcTm{\programTheory}{\tyEnv}{e_1}{\monotype_1 \to \monotype_2}{t_1} \\
            \tcTcTm{\programTheory}{\tyEnv}{e_2}{\monotype_1}{t_2}
           } 
           {\tcTcTm{\programTheory}{\tyEnv}{e_1~e_2}{\monotype_2}{t_1~t_2}}

\inferrule*[right=$(\Rightarrow\hspace{-1mm}I)$]
           {\tcTcCt{\tyEnv}{\constraint}{\fcType} \\
            \tcTcTm{\programTheory\lComma d : \constraint}{\tyEnv}{e}{\qualtype}{t}
           } 
           {\tcTcTm{\programTheory}{\tyEnv}{e}{\constraint \Rightarrow \qualtype}{\lambda (d : \fcType).~t}}

\inferrule*[right=$(\Rightarrow\hspace{-1mm}E)$]
           {\tcTcTm{\programTheory}{\tyEnv}{e}{\constraint \Rightarrow \qualtype}{t_1} \\
            \tcEntailCt{\programTheory}{\tyEnv}{t_2}{\constraint}
           } 
           {\tcTcTm{\programTheory}{\tyEnv}{e}{\qualtype}{t_1~t_2}}

\inferrule*[right=Let]
           { \tcTcTm{\programTheory}{\tyEnv, x : \monotype}{e_1}{\monotype}{t_1} \\
             \tcTcTm{\programTheory}{\tyEnv, x : \monotype}{e_2}{\polytype}{t_2} \\
             \tcTcTy{\tyEnv}{\monotype}{\fcType}
           } 
           { \tcTcTm{\programTheory}{\tyEnv}{\keylet~x = e_1~\keyin~e_2}{\polytype}{\keylet~x : \fcType = t_1~\keyin~t_2} }
\end{mathpar}
\caption{Basic System: Additional Judgments}
\label{fig:basic-system-specification-complete}
\end{figure}

The judgments we omitted in Section~\ref{sec:the-basic-system:specifications}
are given in Figure~\ref{fig:basic-system-specification}, with the
elaboration-related parts highlighted. We briefly describe each below.

\paragraph{Type Well-formedness}

Judgment $\tcTcTyNoShade{\tyEnv}{\polytype}{\fcType}$ captures the
well-formedness and elaboration of types. It checks that under typing
environment $\tyEnv$, type $\polytype$ is well-formed and can be elaborated
into \SystemFC type $\fcType$. Since our system is {\em uni-kinded}, the
relation essentially checks that type $\polytype$ is well-scoped under
environment $\tyEnv$. The only interesting case with respect to elaboration is
that for qualified types, which are elaborated into \SystemFC arrow types.

\paragraph{Constraint Well-formedness}

Constraint well-formedness is given by judgment
$\tcTcCtNoShade{\tyEnv}{\constraint}{\fcType}$ and is equally straightforward.
In essence, a class $\TC$ is elaborated to its corresponding dictionary type
constructor $\tyCon_\TC$.

\paragraph{Term Typing}

Typing and elaboration for terms is captured in judgment
$\tcTcTmNoShade{\programTheory}{\tyEnv}{e}{\polytype}{t}$. Most of the rules
are standard for \HM-based systems. The only interesting rules that relate to
type classes are Rules~\textsc{$(\Rightarrow\hspace{-1mm}I)$}
and~\textsc{$(\Rightarrow\hspace{-1mm}E)$}, which capture qualification
introduction and elimination, respectively. Specifically the latter, which
reflects the elimination in the elaborated term via an explicit dictionary
application, as provided by the constraint entailment relation.

\begin{figure*}
\small
\begin{minipage}{\textwidth}
\begin{flushleft}
\namedRuleform{\fcTcCo{\fcEnv}{\coercion}{\fcEqCt}}
              {Coercion Typing}
\end{flushleft}
\begin{mathpar}
  \inferrule*[right=CoVar]
             {(\covar : \fcEqCt) \in \fcEnv}
             {\fcTcCo{\fcEnv}{\covar}{\fcEqCt}}

  \inferrule*[right=CoAx]
             {(g~\as : \fcType_1 \sim \fcType_2) \in \fcEnv \\
              \overline{\fcTcTy{\fcEnv}{\fcType}}
             } 
             {\fcTcCo{\fcEnv}
                     {g~\fcTypes}
                     {[\fcTypes/\as]\fcType_1 \sim [\fcTypes/\as]\fcType_2}}

  \inferrule*[right=CoRefl]
             {\fcTcTy{\fcEnv}{\fcType}}
             {\fcTcCo{\fcEnv}{\frefl{\fcType}}{\fcType \sim \fcType}}

  \inferrule*[right=CoSym]
             {\fcTcCo{\fcEnv}{\coercion}      {\fcType_1 \sim \fcType_2}}
             {\fcTcCo{\fcEnv}{\fsym{\coercion}}{\fcType_2 \sim \fcType_1}}

  \inferrule*[right=CoTrans]
             {\fcTcCo{\fcEnv}{\coercion_1}{\fcType_1 \sim \fcType_2} \\
              \fcTcCo{\fcEnv}{\coercion_2}{\fcType_2 \sim \fcType_3}
             } 
             {\fcTcCo{\fcEnv}{\fcomp{\coercion_1}{\coercion_2}}{\fcType_1 \sim \fcType_3}}

  \inferrule*[right=CoApp]
             {\fcTcTy{\fcEnv}{\fcType_1~\fcType_3} \\
              \fcTcCo{\fcEnv}{\coercion_1}{\fcType_1 \sim \fcType_2} \\
              \fcTcCo{\fcEnv}{\coercion_2}{\fcType_3 \sim \fcType_4}
             } 
             {\fcTcCo{\fcEnv}{\coercion_1~\coercion_2}{\fcType_1~\fcType_3 \sim \fcType_2~\fcType_4}}

  \inferrule*[right=CoL]
             {\fcTcCo{\fcEnv}{\coercion}{\fcType_1~\fcType_2 \sim \fcType_3~\fcType_4}}
             {\fcTcCo{\fcEnv}{\fleft{\coercion}}{\fcType_1 \sim \fcType_3}}

  \inferrule*[right=CoR]
             {\fcTcCo{\fcEnv}{\coercion}{\fcType_1~\fcType_2 \sim \fcType_3~\fcType_4}}
             {\fcTcCo{\fcEnv}{\fright{\coercion}}{\fcType_2 \sim \fcType_4}}

  \inferrule*[right=CoFam]
             {\tyFam_n~\text{defined} \\
              \overline{\fcTcCo{\fcEnv}{\coercion}{\fcType_1 \sim \fcType_2}}^n \\
              \overline{\fcTcTy{\fcEnv}{\fcType_1}}^n
             } 
             {\fcTcCo{\fcEnv}{\tyFamApp{\tyFam_n}{\overline{\coercion}^n}}{\tyFamApp{\tyFam}{\overline{\fcType_1}^n} \sim \tyFamApp{\tyFam}{\overline{\fcType_2}^n}}}

  \inferrule*[right=CoAll]
             {\fcTcCo{\fcEnv, a}{\coercion}{\fcType_1 \sim \fcType_2} \\
              \fcTcTy{\fcEnv, a}{\fcType_1}
             } 
             {\fcTcCo{\fcEnv}{\blueforall a.~\coercion}{\blueforall a.~\fcType_1 \sim \blueforall a.~\fcType_2}}

  \inferrule*[right=CoIns]
             {\fcTcCo{\fcEnv}{\coercion_1}{\blueforall a.~\fcType_1 \sim \blueforall a.~\fcType_2} \\
              \fcTcCo{\fcEnv}{\coercion_2}{\fcType_3 \sim \fcType_4} \\
              \fcTcTy{\fcEnv}{\fcType_3}
             } 
             {\fcTcCo{\fcEnv}{\finst{\coercion_1}{\coercion_2}}{[\fcType_3/a]\fcType_1 \sim [\fcType_4/a]\fcType_2}}

  \inferrule*[right=CoQual]
             {\fcTcPr{\fcEnv}{\fcEqCt} \\
              \fcTcCo{\fcEnv}{\coercion}{\fcType_1 \sim \fcType_2}
             } 
             {\fcTcCo{\fcEnv}{\fcEqCt \Rightarrow \coercion}{(\fcEqCt \Rightarrow \fcType_1) \sim (\fcEqCt \Rightarrow \fcType_2)}}

  \inferrule*[right=CoQInst]
             {\fcTcCo{\fcEnv}{\coercion_1}{(\fcEqCt \Rightarrow \fcType_1) \sim (\fcEqCt \Rightarrow \fcType_2)} \\
              \fcTcCo{\fcEnv}{\coercion_2}{\fcEqCt}}
             {\fcTcCo{\fcEnv}{\finstQual{\coercion_1}{\coercion_2}}{\fcType_1 \sim \fcType_2}}
\end{mathpar}

\vspace{2mm}
\begin{flushleft}
\namedRuleform{\fcTcTm{\fcEnv}{t}{\fcType}}
              {Term Typing}
\end{flushleft}
\vspace{-2mm}
\begin{mathpar}
  \inferrule*[right=TmVar]
             { (x : \fcType) \in \fcEnv }
             { \fcTcTm{\fcEnv}{x}{\fcType} }

  \inferrule*[right=TmCon]
             { (\dataCon : \fcType) \in \fcEnv }
             { \fcTcTm{\fcEnv}{\dataCon}{\fcType} }

  \inferrule*[right=$(\forall I)$]
             { \fcTcTm{\fcEnv, a}{t}{\fcType} }
             { \fcTcTm{\fcEnv}{\Lambda a.~t}{\blueforall a.~\fcType} }

  \inferrule*[right=$(\forall E)$]
             { \fcTcTy{\fcEnv}{\fcType} \\\\
               \fcTcTm{\fcEnv}{t}{\blueforall a.~\fcType_1}
             } 
             { \fcTcTm{\fcEnv}{t~\fcType}{[\fcType/a]\fcType_1} }

  \inferrule*[right=$(\to\hspace{-1mm}I)$]
             { \fcTcTy{\fcEnv}{\fcType_1} \\\\
               \fcTcTm{\fcEnv, x : \fcType_1}{t}{\fcType_2}
             } 
             { \fcTcTm{\fcEnv}{\lambda (x : \fcType_1).~t}{\fcType_1 \to \fcType_2} }

  \inferrule*[right=$(\to\hspace{-1mm}E)$]
             { \fcTcTm{\fcEnv}{t_1}{\fcType_1 \to \fcType_2} \\\\
               \fcTcTm{\fcEnv}{t_2}{\fcType_1}
             } 
             { \fcTcTm{\fcEnv}{t_1~t_2}{\fcType_2} }

  \inferrule*[right=$(\Rightarrow\hspace{-1mm}I_{\fcEqCt})$]
             {\fcTcPr{\fcEnv}{\fcEqCt} \\\\
              \fcTcTm{\fcEnv, \covar : \fcEqCt}{t}{\fcType}
             } 
             {\fcTcTm{\fcEnv}{\Lambda (\covar : \fcEqCt).~t}{\fcEqCt \Rightarrow \fcType}}

  \inferrule*[right=$(\Rightarrow\hspace{-1mm}E_{\fcEqCt})$]
             {\fcTcTm{\fcEnv}{t}{\fcEqCt \Rightarrow \fcType} \\\\
              \fcTcCo{\fcEnv}{\coercion}{\fcEqCt}
             } 
             {\fcTcTm{\fcEnv}{t~\coercion}{\fcType}}

  \inferrule*[right=TmCast]
             {\fcTcTm{\fcEnv}{t}{\fcType_1} \\\\
              \fcTcCo{\fcEnv}{\coercion}{\fcType_1 \sim \fcType_2}
             } 
             {\fcTcTm{\fcEnv}{\fcCast{t}{\coercion}}{\fcType_2}}

  \inferrule*[right=TmCase]
             { \fcTcTm{\fcEnv}{t_1}{\fcType_1} \\
               \fcTcPat{\fcEnv}{\overline{\fcPat \to t_2}}{\overline{\fcType_1 \to \fcType_2}}
             } 
             { \fcTcTm{\fcEnv}{(\keycase~t_1~\keyof~\overline{\fcPat \to t_2})}{\fcType_2} }

  \inferrule*[right=TmLet]
             { \fcTcTm{\fcEnv, x : \fcType}{t_1}{\fcType} \\
               \fcTcTm{\fcEnv, x : \fcType}{t_2}{\fcType_2}
             } 
             { \fcTcTm{\fcEnv}{(\keylet~x : \fcType = t_1~\keyin~t_2)}{\fcType_2} }
\end{mathpar}

\vspace{2mm}
\begin{flushleft}
\namedRuleform{\fcTcTy{\fcEnv}{\fcType}}
              {Type Well-formedness}
\end{flushleft}
\vspace{-2mm}
\begin{mathpar}
  \inferrule*[right=TyVar]
             { a \in \fcEnv }
             { \fcTcTy{\fcEnv}{a} }

  \inferrule*[right=TyCon]
             { \tyCon \in \fcEnv }
             { \fcTcTy{\fcEnv}{\tyCon} }

  \inferrule*[right=TyAbs]
             { \fcTcTy{\fcEnv, a}{\fcType} }
             { \fcTcTy{\fcEnv}{\blueforall a.~\fcType} }

  \inferrule*[right=TyApp]
             { \fcTcTy{\fcEnv}{\fcType_1} \\\\
               \fcTcTy{\fcEnv}{\fcType_2}
             } 
             { \fcTcTy{\fcEnv}{\fcType_1~\fcType_2} }

  \inferrule*[right=TyFam]
             {\tyFam_n \in \fcEnv \\\\
              \fcTcTy{\fcEnv}{\fcTypes}
             } 
             {\fcTcTy{\fcEnv}{\tyFamApp{\tyFam_n}{\fcTypes}}}

  \inferrule*[right=TyQual]
             {\fcTcPr{\fcEnv}{\fcEqCt} \\\\
              \fcTcTy{\fcEnv}{\fcType}
             } 
             {\fcTcTy{\fcEnv}{\fcEqCt \Rightarrow \fcType}}
\end{mathpar}

\vspace{2mm}
\begin{minipage}{0.70\textwidth}
\begin{flushleft}
\namedRuleform{\fcTcPat{\fcEnv}{\fcPat \to t}{\fcType_1 \to \fcType_2}}
              {Pattern Typing}
\end{flushleft}
\begin{mathpar}
  \inferrule*[right=Pat]
             { (\dataCon : \blueforall \as \bs'.~\fcEqCs \Rightarrow \fcTypes \to \tyCon~\as) \in \fcEnv \\
               \tySubst = [\fcTypes_a/\as, \bs/\bs'] \\
               \fcTcTm{\fcEnv, \bs, (\overline{\covar : \tySubst(\fcEqCt)}), (\overline{x : \tySubst(\fcType)})}{t}{\fcType_2}
             } 
             {\fcTcPat{\fcEnv}{\dataCon~\bs~(\overline{\covar : \tySubst(\fcEqCt)})~(\overline{x : \tySubst(\fcType)}) \to t}{\tyCon~\fcTypes_a \to \fcType_2}}
\end{mathpar}
\end{minipage}
\begin{minipage}{0.29\textwidth}
\begin{flushleft}
\namedRuleform{\fcTcPr{\fcEnv}{\fcEqCt}}
              {Proposition Well-formedness}
\end{flushleft}
\begin{mathpar}
  \inferrule*[right=Prop]
             {\fcTcTy{\fcEnv}{\fcType_1} \\
              \fcTcTy{\fcEnv}{\fcType_2}
             } 
             {\fcTcPr{\fcEnv}{\fcType_1 \sim \fcType_2}}
\end{mathpar}
\end{minipage}

\vspace{2mm}
\begin{flushleft}
\namedRuleform{\fcTcDecl{\fcEnv_1}{\decl}{\fcEnv_2}}
              {Declaration Typing}
\end{flushleft}
\begin{mathpar}
\inferrule*[right=Data]
           { \fcType_{\dataCon_i} \equiv \blueforall \as \bs_i.~\fcEqCs_i \Rightarrow \fcTypes_i \to \tyCon~\as \\
             \fcTcTy{\fcEnv}{\fcType_{\dataCon_i}}
           } 
           { \fcTcDecl{\fcEnv}{(\keydata~\tyCon~\as~\keywhere~\{~\overline{\dataCon : \fcType_\dataCon}~\})}{[\tyCon, \overline{\dataCon : \fcType_\dataCon}]} }

\inferrule*[right=Family]
           { }
           { \fcTcDecl{\fcEnv}{(\keytype~\tyFamApp{\tyFam}{\as^n})}{[\tyFam_n]} }

\inferrule*[right=Axiom]
           { \fcTcTy{\fcEnv, \as}{\tyPat_i} \\
             \fcTcTy{\fcEnv, \as}{\fcType}
           } 
           { \fcTcDecl{\fcEnv}{(\keyaxiom~g~\as : \tyFamApp{\tyFam}{\tyPats} \sim \fcType)}{[g~\as : \tyFamApp{\tyFam}{\tyPats} \sim \fcType]} }

\inferrule*[right=Value]
           { \fcTcTm{\fcEnv, x : \fcType}{t}{\fcType} }
           { \fcTcDecl{\fcEnv}{(\keylet~x : \fcType = t)}{[x : \fcType]} }
\end{mathpar}
\end{minipage}
\caption{\SystemFC Typing}
\label{fig:class-background:systemfc-coercion-typing}
\end{figure*}

\subsection{Type Inference and Elaboration Algorithm}

We now present and briefly discuss the judgments we omitted in
Section~\ref{sec:the-basic-system:algorithms}.

\subsubsection{Constraint Generation}

\begin{figure}
\small
\begin{flushleft}
\namedRuleform{\tcElabTm{\tyEnv}{e}{\monotype}{t}{\annConstraintSet}{\equalities}}
              {Constraint Generation}
\end{flushleft}
\begin{mathpar}
\inferrule*[right=Var]
           { (x : \blueforall \as.~\overline{\constraint} \Rightarrow \monotype) \in \tyEnv \\
             \tySubst = [\bs/\as]
           } 
           { \tcElabTm{\tyEnv}{x}{\tySubst(\monotype)}
                     {x~\bs~\ds}
                     {\overline{(\highlight{d :} \tySubst(\constraint))}}
                     {\bullet}
           }

\inferrule*[right=Let]
           { \tcElabTm{\tyEnv, x : a}{e_1}{\monotype_1}{t_1}{\annConstraintSet_1}{\equalities_1} \\
             \tcElabTm{\tyEnv, x : \monotype_1}{e_2}{\monotype_2}{t_2}{\annConstraintSet_2}{\equalities_2} \\
             \highlight{t = \keylet~x : \tcElabTy{\monotype_1} = t_1~\keyin~t_2}
           }
           { \tcElabTm{\tyEnv}{\keylet~x=e_1~\keyin~e_2}{\monotype_2}
                     {t} 
                     {(\annConstraintSet_1, \annConstraintSet_2)}
                     {(\equalities_1, \equalities_2, a \sim \monotype_1)} 
           }

\inferrule*[right=$(\rightarrow\hspace{-1mm}I)$]
           { \tcElabTm{\tyEnv, x : a}{e}{\monotype}{t}{\annConstraintSet}{\equalities} }
           { \tcElabTm{\tyEnv}{\lambda x.~e}{a \to \monotype}{\lambda (x : a).~t}{\annConstraintSet}{\equalities} }

\inferrule*[right=$(\rightarrow\hspace{-1mm}E)$]
           { \tcElabTm{\tyEnv}{e_1}{\monotype_1}{t_1}{\annConstraintSet_1}{\equalities_1} \\
             \tcElabTm{\tyEnv}{e_2}{\monotype_2}{t_2}{\annConstraintSet_2}{\equalities_2}
           }
           { \tcElabTm{\tyEnv}{e_1~e_2}{a}{t_1~t_2}{(\annConstraintSet_1, \annConstraintSet_2)}
                     {(\equalities_1, \equalities_2, \monotype_1 \sim \monotype_2 \to a)}
           }
\end{mathpar}
\caption{Term Elaboration and Constraint Generation}
\label{fig:class-background:constraint-gen-elaboration}
\end{figure}

Constraint generation with elaboration for terms takes the form
$\tcElabTmNoShade{\tyEnv}{e}{\monotype}{t}{\annConstraintSet}{\equalities}$ and
is presented in Figure~\ref{fig:class-background:constraint-gen-elaboration}.
Given a typing environment $\tyEnv$ and a source expression $e$, we infer a
monotype $\monotype$ for $e$ and generate wanted constraints
$\annConstraintSet$ and $\equalities$. At the same time, we elaborate $e$ into
\SystemFC term $t$.

Rule~\textsc{Var} handles term variables. The polymorphic type $\blueforall
\as.~\overline{\constraint} \Rightarrow \monotype$ of a term variable $x$ is
instantiated with fresh {\em unification} variables $\bs$, and constraints
$\overline{\constraint}$ are introduced as wanted constraints, instantiated
likewise. In the elaborated term instantiation becomes explicit via type
application. Similarly, the source-level elimination of constraints
$\overline{\constraint}$ amounts to term-level application in \SystemFC.
Arguments $\ds$ capture the yet-unknown dictionaries, evidence for the wanted
constraints $\overline{\constraint}$.

Rule~\textsc{Let} handles (possibly recursive) monomorphic let-bindings.  After
assigning a fresh unification variable $a$ to the term variable $x$, we infer
types for both $e_1$ and $e_2$. We choose not to perform
let-generalization,\footnote{This is a mere simplification for presentational
purposes; bidirectional instances are orthogonal to let-generalization.} so
Rule~\textsc{Let} does not make a distinction between constraints generated by
$e_1$ or $e_2$; they are both part of the result.  Finally, we record that the
(monomorphic) type of $x$ is equal to the type of the term it is bound to: $a
\sim \monotype_1$.

Rule~\textsc{TmAbs} is straightforward: we generate a fresh type variable for
the argument $x$, and collect constraints generated from typing the body.
Rule~\textsc{TmApp} combines the wanted constraints from both subterms, and
records that the application is well-formed via equality $(\monotype_1 \sim
\monotype_2 \to a)$.

%
%

\subsubsection{Elaboration of Types and Constraints}

Elaboration of types and constraints is given by functions
$\tcElabTy{\polytype} = \fcType$ and $\tcElabCt{\constraint} = \fcType$,
respectively. The former is given by the following clauses
{\small
\[
\begin{array}{l@{\hspace{1mm}}c@{\hspace{1mm}}l}
  \tcElabTy{a}                                 & = & a                                                 \\
  \tcElabTy{\monotype_1 \to \monotype_2}       & = & \tcElabTy{\monotype_1} \to \tcElabTy{\monotype_2} \\
  \tcElabTy{\constraint \Rightarrow \qualtype} & = & \tcElabCt{\constraint} \to \tcElabTy{\qualtype}   \\
  \tcElabTy{\blueforall a.~\polytype}              & = & \blueforall a.~\tcElabTy{\polytype}                   \\
\end{array}
\]
}
and the latter by a single clause:
{\small
\[
\begin{array}{l@{\hspace{1mm}}c@{\hspace{1mm}}l}
  \tcElabCt{\TC~\monotype} & = & \tyCon_\TC~\tcElabTy{\monotype} \\
\end{array}
\]
}
Both functions are straightforward; the only interesting aspect is the
elaboration of qualified types into \SystemFC arrow types. Class constraints
$\TC~\monotype$ are elaborated into dictionary types
$\tyCon_\TC~\tcElabTy{\monotype}$, where type constructor $\tyCon_\TC$ is the
\SystemFC representation of class $\TC$.

\subsubsection{Hindley-Damas-Milner Unification}

The standard \HM type unification algorithm we omitted in
Section~\ref{sec:the-basic-system:algorithms} ($\tcUnify{\as}{\equalities} =
\tySubst_\bot$) is given by the following rules:

{\small
\[
\begin{array}{@{\hspace{0mm}}l@{\hspace{2mm}}c@{\hspace{2mm}}l@{\hspace{0mm}}}
  \tcUnify{\as}{\bullet}                       & = & \bullet \\
  \tcUnify{\as}{\equalities, b \sim b}         & = & \tcUnify{\as}{\equalities} \\
  \tcUnify{\as}{\equalities, b \sim \monotype} & = & \tcUnify{\as}{\tySubst(\equalities)} \cdot \tySubst \\
    \multicolumn{3}{@{\hspace{0mm}}l}{\qquad\text{where } b \notin \as \land b \notin \fv{\monotype} \land \tySubst = [\monotype/b]} \\
  \tcUnify{\as}{\equalities, \monotype \sim b} & = & \tcUnify{\as}{\tySubst(\equalities)} \cdot \tySubst \\
    \multicolumn{3}{@{\hspace{0mm}}l}{\qquad\text{where } b \notin \as \land b \notin \fv{\monotype} \land \tySubst = [\monotype/b]} \\
  \tcUnify{\as}{\equalities, (\monotype_1 \to \monotype_2) \sim (\monotype_3 \to \monotype_4)} & = &
    \tcUnify{\as}{\equalities, \monotype_1 \sim \monotype_3, \monotype_2 \sim \monotype_4} \\
\end{array}
\]
}

\section{\SystemFC Specification}
\label{appendix:remaining-systemfc}

Typing for the dialect of \SystemFC we target in this work is presented in
Figure~\ref{fig:class-background:systemfc-coercion-typing}. All judgments are
parameterized over target typing environments $\fcEnv$, defined as follows:
\[
\fcEnv ::=  \bullet
       \mid \fcEnv, a
       \mid \fcEnv, x : \fcType
       \mid \fcEnv, \tyCon
       \mid \fcEnv, \dataCon : \fcType
       \mid \fcEnv, \covar : \fcEqCt
       \mid \fcEnv, g~\as : \fcEqCt
\]
For a more detailed description of \SystemFC, we urge the reader to consult its
original publication by~\citet{systemfc}.

\begin{acks}                            

We would like to thank Steven Keuchel, and the anonymous reviewers of Haskell
Symposium 2018 and 2019 for careful reading and insightful comments.
This research was supported by the Flemish Fund for Scientific
Research (FWO).
\end{acks}

\bibliography{references}


\begin{thebibliography}{38}


\ifx \showCODEN    \undefined \def \showCODEN     #1{\unskip}     \fi
\ifx \showDOI      \undefined \def \showDOI       #1{#1}\fi
\ifx \showISBNx    \undefined \def \showISBNx     #1{\unskip}     \fi
\ifx \showISBNxiii \undefined \def \showISBNxiii  #1{\unskip}     \fi
\ifx \showISSN     \undefined \def \showISSN      #1{\unskip}     \fi
\ifx \showLCCN     \undefined \def \showLCCN      #1{\unskip}     \fi
\ifx \shownote     \undefined \def \shownote      #1{#1}          \fi
\ifx \showarticletitle \undefined \def \showarticletitle #1{#1}   \fi
\ifx \showURL      \undefined \def \showURL       {\relax}        \fi
\providecommand\bibfield[2]{#2}
\providecommand\bibinfo[2]{#2}
\providecommand\natexlab[1]{#1}
\providecommand\showeprint[2][]{arXiv:#2}

\bibitem[\protect\citeauthoryear{Barendregt}{Barendregt}{1981}]%
        {barendregt}
\bibfield{author}{\bibinfo{person}{Henk Barendregt}.}
  \bibinfo{year}{1981}\natexlab{}.
\newblock \bibinfo{booktitle}{\emph{The Lambda Calculus: its Syntax and
  Semantics, volume 103 of Studies in Logic and the Foundations of
  Mathematics}}.
\newblock


\bibitem[\protect\citeauthoryear{Bottu, Karachalias, Schrijvers, Oliveira, and
  Wadler}{Bottu et~al\mbox{.}}{2017}]%
        {quantcs}
\bibfield{author}{\bibinfo{person}{Gert-Jan Bottu}, \bibinfo{person}{Georgios
  Karachalias}, \bibinfo{person}{Tom Schrijvers}, \bibinfo{person}{Bruno C.
  d.~S. Oliveira}, {and} \bibinfo{person}{Philip Wadler}.}
  \bibinfo{year}{2017}\natexlab{}.
\newblock \showarticletitle{Quantified Class Constraints}. In
  \bibinfo{booktitle}{\emph{Haskell 2017}}.
\newblock
\showISBNx{978-1-4503-5182-9}


\bibitem[\protect\citeauthoryear{Bottu, Xie, Marntirosian, and
  Schrijvers}{Bottu et~al\mbox{.}}{2019}]%
        {CoherenceProof}
\bibfield{author}{\bibinfo{person}{Gert-Jan Bottu}, \bibinfo{person}{Ningning
  Xie}, \bibinfo{person}{Klara Marntirosian}, {and} \bibinfo{person}{Tom
  Schrijvers}.} \bibinfo{year}{2019}\natexlab{}.
\newblock \showarticletitle{Coherence of Type Class Resolution}.
\newblock \bibinfo{journal}{\emph{Proc. ACM Program. Lang.}}
  (\bibinfo{year}{2019}).
\newblock
\newblock
\shownote{Accepted.}


\bibitem[\protect\citeauthoryear{Chakravarty, Keller, and Jones}{Chakravarty
  et~al\mbox{.}}{2005a}]%
        {AssociatedTypeSynonyms}
\bibfield{author}{\bibinfo{person}{Manuel M.~T. Chakravarty},
  \bibinfo{person}{Gabriele Keller}, {and} \bibinfo{person}{Simon~Peyton
  Jones}.} \bibinfo{year}{2005}\natexlab{a}.
\newblock \showarticletitle{Associated Type Synonyms}.
\newblock \bibinfo{journal}{\emph{SIGPLAN Not.}} \bibinfo{volume}{40},
  \bibinfo{number}{9} (\bibinfo{year}{2005}), \bibinfo{pages}{241--253}.
\newblock
\showISSN{0362-1340}


\bibitem[\protect\citeauthoryear{Chakravarty, Keller, Jones, and
  Marlow}{Chakravarty et~al\mbox{.}}{2005b}]%
        {DataFamilies}
\bibfield{author}{\bibinfo{person}{Manuel M.~T. Chakravarty},
  \bibinfo{person}{Gabriele Keller}, \bibinfo{person}{Simon~Peyton Jones},
  {and} \bibinfo{person}{Simon Marlow}.} \bibinfo{year}{2005}\natexlab{b}.
\newblock \showarticletitle{Associated Types with Class}. In
  \bibinfo{booktitle}{\emph{POPL '05}}. \bibinfo{publisher}{ACM},
  \bibinfo{pages}{1--13}.
\newblock
\showISBNx{1-58113-830-X}


\bibitem[\protect\citeauthoryear{Damas and Milner}{Damas and Milner}{1982}]%
        {DamasMilner}
\bibfield{author}{\bibinfo{person}{Luis Damas} {and} \bibinfo{person}{Robin
  Milner}.} \bibinfo{year}{1982}\natexlab{}.
\newblock \showarticletitle{Principal Type-schemes for Functional Programs}. In
  \bibinfo{booktitle}{\emph{POPL '82}}. \bibinfo{publisher}{ACM},
  \bibinfo{pages}{207--212}.
\newblock
\showISBNx{0-89791-065-6}


\bibitem[\protect\citeauthoryear{Eisenberg, Vytiniotis, Peyton~Jones, and
  Weirich}{Eisenberg et~al\mbox{.}}{2014}]%
        {closedtf}
\bibfield{author}{\bibinfo{person}{Richard~A. Eisenberg},
  \bibinfo{person}{Dimitrios Vytiniotis}, \bibinfo{person}{Simon Peyton~Jones},
  {and} \bibinfo{person}{Stephanie Weirich}.} \bibinfo{year}{2014}\natexlab{}.
\newblock \showarticletitle{Closed Type Families with Overlapping Equations}.
  In \bibinfo{booktitle}{\emph{POPL '14}}.
\newblock
\showISBNx{978-1-4503-2544-8}


\bibitem[\protect\citeauthoryear{Girard}{Girard}{1972}]%
        {girardthesis}
\bibfield{author}{\bibinfo{person}{Jean-Yves Girard}.}
  \bibinfo{year}{1972}\natexlab{}.
\newblock \emph{\bibinfo{title}{Interpr{\'e}tation fonctionelle et
  {\'e}limination des coupures de l'arithm{\'e}tique d'ordre sup{\'e}rieur}}.
\newblock \bibinfo{thesistype}{Ph.D. Dissertation}.
\newblock


\bibitem[\protect\citeauthoryear{Gregor, J\"{a}rvi, Siek, Stroustrup, Dos~Reis,
  and Lumsdaine}{Gregor et~al\mbox{.}}{2006}]%
        {concepts-article}
\bibfield{author}{\bibinfo{person}{Douglas Gregor}, \bibinfo{person}{Jaakko
  J\"{a}rvi}, \bibinfo{person}{Jeremy Siek}, \bibinfo{person}{Bjarne
  Stroustrup}, \bibinfo{person}{Gabriel Dos~Reis}, {and}
  \bibinfo{person}{Andrew Lumsdaine}.} \bibinfo{year}{2006}\natexlab{}.
\newblock \showarticletitle{Concepts: Linguistic Support for Generic
  Programming in C++}.
\newblock \bibinfo{journal}{\emph{SIGPLAN Not.}} \bibinfo{volume}{41},
  \bibinfo{number}{10} (\bibinfo{year}{2006}), \bibinfo{pages}{291--310}.
\newblock
\showISSN{0362-1340}


\bibitem[\protect\citeauthoryear{Hall, Hammond, Peyton~Jones, and Wadler}{Hall
  et~al\mbox{.}}{1996}]%
        {ClassElaboration}
\bibfield{author}{\bibinfo{person}{Cordelia~V. Hall}, \bibinfo{person}{Kevin
  Hammond}, \bibinfo{person}{Simon~L. Peyton~Jones}, {and}
  \bibinfo{person}{Philip~L. Wadler}.} \bibinfo{year}{1996}\natexlab{}.
\newblock \showarticletitle{Type Classes in Haskell}.
\newblock \bibinfo{journal}{\emph{TOPLAS}} \bibinfo{volume}{18},
  \bibinfo{number}{2} (\bibinfo{date}{March} \bibinfo{year}{1996}).
\newblock
\showISSN{0164-0925}


\bibitem[\protect\citeauthoryear{Hallgren}{Hallgren}{2000}]%
        {FunWithFDs}
\bibfield{author}{\bibinfo{person}{Thomas Hallgren}.}
  \bibinfo{year}{2000}\natexlab{}.
\newblock \showarticletitle{Fun with Functional Dependencies}. In
  \bibinfo{booktitle}{\emph{Proc. of the Joint CS/CE Winter Meeting}}.
\newblock


\bibitem[\protect\citeauthoryear{Henderson, Conway, Somogyi, Jeffery, Schachte,
  Taylor, and Speirs}{Henderson et~al\mbox{.}}{1996}]%
        {mercury}
\bibfield{author}{\bibinfo{person}{Fergus Henderson}, \bibinfo{person}{Thomas
  Conway}, \bibinfo{person}{Zoltan Somogyi}, \bibinfo{person}{David Jeffery},
  \bibinfo{person}{Peter Schachte}, \bibinfo{person}{Simon Taylor}, {and}
  \bibinfo{person}{Chris Speirs}.} \bibinfo{year}{1996}\natexlab{}.
\newblock \bibinfo{booktitle}{\emph{The Mercury Language Reference Manual}}.
\newblock \bibinfo{type}{{T}echnical {R}eport}.
\newblock


\bibitem[\protect\citeauthoryear{Henglein}{Henglein}{1993}]%
        {UndecidablePolymorphicRecursion}
\bibfield{author}{\bibinfo{person}{Fritz Henglein}.}
  \bibinfo{year}{1993}\natexlab{}.
\newblock \showarticletitle{Type Inference with Polymorphic Recursion}.
\newblock \bibinfo{journal}{\emph{ACM Trans. Program. Lang. Syst.}}
  \bibinfo{volume}{15}, \bibinfo{number}{2} (\bibinfo{date}{April}
  \bibinfo{year}{1993}), \bibinfo{pages}{253--289}.
\newblock
\showISSN{0164-0925}


\bibitem[\protect\citeauthoryear{Hindley}{Hindley}{1969}]%
        {hindley}
\bibfield{author}{\bibinfo{person}{R. Hindley}.}
  \bibinfo{year}{1969}\natexlab{}.
\newblock \showarticletitle{{The Principal Type-Scheme of an Object in
  Combinatory Logic}}.
\newblock \bibinfo{journal}{\emph{Trans. Amer. Math. Soc.}}
  \bibinfo{volume}{146} (\bibinfo{year}{1969}), \bibinfo{pages}{29--60}.
\newblock
\showISSN{00029947}


\bibitem[\protect\citeauthoryear{Johann and Ghani}{Johann and Ghani}{2008}]%
        {FoundationsGADTs}
\bibfield{author}{\bibinfo{person}{Patricia Johann} {and} \bibinfo{person}{Neil
  Ghani}.} \bibinfo{year}{2008}\natexlab{}.
\newblock \showarticletitle{Foundations for Structured Programming with GADTs}.
\newblock \bibinfo{journal}{\emph{SIGPLAN Not.}} \bibinfo{volume}{43},
  \bibinfo{number}{1} (\bibinfo{date}{Jan.} \bibinfo{year}{2008}),
  \bibinfo{pages}{297--308}.
\newblock
\showISSN{0362-1340}


\bibitem[\protect\citeauthoryear{Jones}{Jones}{1992}]%
        {qualtypes}
\bibfield{author}{\bibinfo{person}{Mark~P. Jones}.}
  \bibinfo{year}{1992}\natexlab{}.
\newblock \showarticletitle{A theory of qualified types}.
\newblock In \bibinfo{booktitle}{\emph{ESOP '92}},
  \bibfield{editor}{\bibinfo{person}{Bernd Krieg-Br\"{u}ckner}} (Ed.).
  \bibinfo{series}{LNCS}, Vol.~\bibinfo{volume}{582}.
  \bibinfo{publisher}{Springer Berlin Heidelberg}, \bibinfo{pages}{287--306}.
\newblock


\bibitem[\protect\citeauthoryear{Jones}{Jones}{1995a}]%
        {JonesThesis}
\bibfield{author}{\bibinfo{person}{Mark~P. Jones}.}
  \bibinfo{year}{1995}\natexlab{a}.
\newblock \bibinfo{booktitle}{\emph{Qualified Types: Theory and Practice}}.
\newblock \bibinfo{publisher}{Cambridge University Press}.
\newblock
\showISBNx{0-521-47253-9}


\bibitem[\protect\citeauthoryear{Jones}{Jones}{1995b}]%
        {JonesImprovement}
\bibfield{author}{\bibinfo{person}{Mark~P. Jones}.}
  \bibinfo{year}{1995}\natexlab{b}.
\newblock \showarticletitle{Simplifying and Improving Qualified Types}. In
  \bibinfo{booktitle}{\emph{FPCA '95}}. \bibinfo{publisher}{ACM},
  \bibinfo{pages}{160--169}.
\newblock


\bibitem[\protect\citeauthoryear{Jones}{Jones}{2000}]%
        {fundeps-original}
\bibfield{author}{\bibinfo{person}{Mark~P. Jones}.}
  \bibinfo{year}{2000}\natexlab{}.
\newblock \showarticletitle{Type Classes with Functional Dependencies}.
\newblock In \bibinfo{booktitle}{\emph{Programming Languages and Systems}}.
  \bibinfo{series}{LNCS}, Vol.~\bibinfo{volume}{1782}.
  \bibinfo{publisher}{Springer}.
\newblock


\bibitem[\protect\citeauthoryear{Karachalias and Schrijvers}{Karachalias and
  Schrijvers}{2017}]%
        {fundeps-core}
\bibfield{author}{\bibinfo{person}{Georgios Karachalias} {and}
  \bibinfo{person}{Tom Schrijvers}.} \bibinfo{year}{2017}\natexlab{}.
\newblock \showarticletitle{Elaboration on Functional Dependencies: Functional
  Dependencies Are Dead, Long Live Functional Dependencies!}
\newblock \bibinfo{journal}{\emph{SIGPLAN Not.}} \bibinfo{volume}{52},
  \bibinfo{number}{10} (\bibinfo{date}{Sept.} \bibinfo{year}{2017}),
  \bibinfo{pages}{133--147}.
\newblock
\showISSN{0362-1340}


\bibitem[\protect\citeauthoryear{Kowalski}{Kowalski}{1974}]%
        {sld-resolution}
\bibfield{author}{\bibinfo{person}{Robert Kowalski}.}
  \bibinfo{year}{1974}\natexlab{}.
\newblock \showarticletitle{Predicate Logic as Programming Language}. In
  \bibinfo{booktitle}{\emph{Proceedings of IFIP '74}}. \bibinfo{address}{North
  Holland}, \bibinfo{pages}{569--574}.
\newblock


\bibitem[\protect\citeauthoryear{\mbox{The Coq development team}}{\mbox{The Coq
  development team}}{2004}]%
        {Coq:manual}
\bibfield{author}{\bibinfo{person}{\mbox{The Coq development team}}.}
  \bibinfo{year}{2004}\natexlab{}.
\newblock \bibinfo{booktitle}{\emph{The Coq proof assistant reference manual}}.
\newblock LogiCal Project.
\newblock
\urldef\tempurl%
\url{http://coq.inria.fr}
\showURL{%
\tempurl}
\newblock
\shownote{Version 8.0.}


\bibitem[\protect\citeauthoryear{Morris}{Morris}{2014}]%
        {SimpleSemanticsForHaskellOverloading}
\bibfield{author}{\bibinfo{person}{J.~Garrett Morris}.}
  \bibinfo{year}{2014}\natexlab{}.
\newblock \showarticletitle{A Simple Semantics for Haskell Overloading}.
\newblock \bibinfo{journal}{\emph{SIGPLAN Not.}} \bibinfo{volume}{49},
  \bibinfo{number}{12} (\bibinfo{year}{2014}), \bibinfo{pages}{107--118}.
\newblock
\showISSN{0362-1340}


\bibitem[\protect\citeauthoryear{Morris and Eisenberg}{Morris and
  Eisenberg}{2017}]%
        {constrained-type-families}
\bibfield{author}{\bibinfo{person}{J.~Garrett Morris} {and}
  \bibinfo{person}{Richard~A. Eisenberg}.} \bibinfo{year}{2017}\natexlab{}.
\newblock \showarticletitle{Constrained Type Families}. In
  \bibinfo{booktitle}{\emph{ICFP '17}}. \bibinfo{publisher}{ACM}.
\newblock


\bibitem[\protect\citeauthoryear{Morris and Jones}{Morris and Jones}{2010}]%
        {InstanceChains}
\bibfield{author}{\bibinfo{person}{J.~Garrett Morris} {and}
  \bibinfo{person}{Mark~P. Jones}.} \bibinfo{year}{2010}\natexlab{}.
\newblock \showarticletitle{Instance Chains: Type Class Programming Without
  Overlapping Instances}. In \bibinfo{booktitle}{\emph{ICFP '10}}.
  \bibinfo{publisher}{ACM}.
\newblock


\bibitem[\protect\citeauthoryear{Peano}{Peano}{1889}]%
        {peano-axioms}
\bibfield{author}{\bibinfo{person}{Giuseppe Peano}.}
  \bibinfo{year}{1889}\natexlab{}.
\newblock \bibinfo{booktitle}{\emph{Arithmetices principia: nova methodo
  exposita}}.
\newblock \bibinfo{publisher}{Fratres Bocca}.
\newblock
\urldef\tempurl%
\url{https://books.google.be/books?id=UUFtAAAAMAAJ}
\showURL{%
\tempurl}


\bibitem[\protect\citeauthoryear{Peyton~Jones}{Peyton~Jones}{2003}]%
        {haskell98}
\bibfield{author}{\bibinfo{person}{Simon Peyton~Jones}.}
  \bibinfo{year}{2003}\natexlab{}.
\newblock \bibinfo{booktitle}{\emph{Haskell 98 Language and Libraries: The
  Revised Report}}.
\newblock \bibinfo{publisher}{Cambridge University Press}.
\newblock
\showLCCN{2003273365}


\bibitem[\protect\citeauthoryear{Peyton~Jones, Vytiniotis, Weirich, and
  Washburn}{Peyton~Jones et~al\mbox{.}}{2006}]%
        {PeytonJones06}
\bibfield{author}{\bibinfo{person}{Simon Peyton~Jones},
  \bibinfo{person}{Dimitrios Vytiniotis}, \bibinfo{person}{Stephanie Weirich},
  {and} \bibinfo{person}{Geoffrey Washburn}.} \bibinfo{year}{2006}\natexlab{}.
\newblock \showarticletitle{Simple Unification-based Type Inference for GADTs}.
\newblock \bibinfo{journal}{\emph{SIGPLAN Not.}} \bibinfo{volume}{41},
  \bibinfo{number}{9} (\bibinfo{year}{2006}), \bibinfo{pages}{50--61}.
\newblock
\showISSN{0362-1340}


\bibitem[\protect\citeauthoryear{Ramsey, Dias, and Peyton~Jones}{Ramsey
  et~al\mbox{.}}{2010}]%
        {hoopl}
\bibfield{author}{\bibinfo{person}{Norman Ramsey}, \bibinfo{person}{Jo\~{a}o
  Dias}, {and} \bibinfo{person}{Simon Peyton~Jones}.}
  \bibinfo{year}{2010}\natexlab{}.
\newblock \showarticletitle{Hoopl: A Modular, Reusable Library for Dataflow
  Analysis and Transformation}.
\newblock \bibinfo{journal}{\emph{SIGPLAN Not.}} \bibinfo{volume}{45},
  \bibinfo{number}{11} (\bibinfo{date}{Sept.} \bibinfo{year}{2010}),
  \bibinfo{pages}{121--134}.
\newblock
\showISSN{0362-1340}


\bibitem[\protect\citeauthoryear{Reynolds}{Reynolds}{1974}]%
        {reynolds-systemf-1}
\bibfield{author}{\bibinfo{person}{John~C. Reynolds}.}
  \bibinfo{year}{1974}\natexlab{}.
\newblock \showarticletitle{Towards a Theory of Type Structure}. In
  \bibinfo{booktitle}{\emph{Programming Symposium, Proceedings Colloque Sur La
  Programmation}}. \bibinfo{pages}{408--423}.
\newblock
\showISBNx{3-540-06859-7}


\bibitem[\protect\citeauthoryear{Reynolds}{Reynolds}{1983a}]%
        {reynolds-systemf-2}
\bibfield{author}{\bibinfo{person}{John~C. Reynolds}.}
  \bibinfo{year}{1983}\natexlab{a}.
\newblock \showarticletitle{Types, Abstraction, and Parametric Polymorphism}.
  In \bibinfo{booktitle}{\emph{Information Processing 83}},
  \bibfield{editor}{\bibinfo{person}{R.E.A. Mason}} (Ed.).
  \bibinfo{pages}{513--523}.
\newblock


\bibitem[\protect\citeauthoryear{Reynolds}{Reynolds}{1983b}]%
        {reynolds1983types}
\bibfield{author}{\bibinfo{person}{John~C. Reynolds}.}
  \bibinfo{year}{1983}\natexlab{b}.
\newblock \showarticletitle{Types, abstraction and parametric polymorphism}.
\newblock  (\bibinfo{year}{1983}).
\newblock


\bibitem[\protect\citeauthoryear{Schrijvers, Peyton~Jones, Chakravarty, and
  Sulzmann}{Schrijvers et~al\mbox{.}}{2008}]%
        {typecheckingwithopentf}
\bibfield{author}{\bibinfo{person}{Tom Schrijvers}, \bibinfo{person}{Simon
  Peyton~Jones}, \bibinfo{person}{Manuel Chakravarty}, {and}
  \bibinfo{person}{Martin Sulzmann}.} \bibinfo{year}{2008}\natexlab{}.
\newblock \showarticletitle{Type Checking with Open Type Functions}. In
  \bibinfo{booktitle}{\emph{ICFP '08}}. \bibinfo{publisher}{ACM},
  \bibinfo{pages}{51--62}.
\newblock
\showISBNx{978-1-59593-919-7}


\bibitem[\protect\citeauthoryear{Sulzmann, Chakravarty, Peyton~Jones, and
  Donnelly}{Sulzmann et~al\mbox{.}}{2007a}]%
        {systemfc}
\bibfield{author}{\bibinfo{person}{M. Sulzmann}, \bibinfo{person}{M.~M.~T.
  Chakravarty}, \bibinfo{person}{S. Peyton~Jones}, {and} \bibinfo{person}{K.
  Donnelly}.} \bibinfo{year}{2007}\natexlab{a}.
\newblock \showarticletitle{System {F} with Type Equality Coercions}. In
  \bibinfo{booktitle}{\emph{TLDI '07}}. \bibinfo{publisher}{ACM}.
\newblock


\bibitem[\protect\citeauthoryear{Sulzmann, Duck, Peyton-Jones, and
  Stuckey}{Sulzmann et~al\mbox{.}}{2007b}]%
        {fundeps-chr}
\bibfield{author}{\bibinfo{person}{Martin Sulzmann},
  \bibinfo{person}{Gregory~J. Duck}, \bibinfo{person}{Simon Peyton-Jones},
  {and} \bibinfo{person}{Peter~J. Stuckey}.} \bibinfo{year}{2007}\natexlab{b}.
\newblock \showarticletitle{Understanding Functional Dependencies via
  Constraint Handling Rules}.
\newblock \bibinfo{journal}{\emph{J. Funct. Program.}} \bibinfo{volume}{17},
  \bibinfo{number}{1} (\bibinfo{year}{2007}), \bibinfo{pages}{83--129}.
\newblock


\bibitem[\protect\citeauthoryear{Vytiniotis, Peyton~jones, Schrijvers, and
  Sulzmann}{Vytiniotis et~al\mbox{.}}{2011}]%
        {outsideinx}
\bibfield{author}{\bibinfo{person}{Dimitrios Vytiniotis},
  \bibinfo{person}{Simon Peyton~jones}, \bibinfo{person}{Tom Schrijvers}, {and}
  \bibinfo{person}{Martin Sulzmann}.} \bibinfo{year}{2011}\natexlab{}.
\newblock \showarticletitle{Outsidein(x) Modular Type Inference with Local
  Assumptions}.
\newblock \bibinfo{journal}{\emph{J. Funct. Program.}} \bibinfo{volume}{21},
  \bibinfo{number}{4-5} (\bibinfo{date}{Sept.} \bibinfo{year}{2011}),
  \bibinfo{pages}{333--412}.
\newblock


\bibitem[\protect\citeauthoryear{Wadler and Blott}{Wadler and Blott}{1989}]%
        {adhoc-polymorphism}
\bibfield{author}{\bibinfo{person}{P. Wadler} {and} \bibinfo{person}{S.
  Blott}.} \bibinfo{year}{1989}\natexlab{}.
\newblock \showarticletitle{How to Make Ad-hoc Polymorphism Less Ad Hoc}. In
  \bibinfo{booktitle}{\emph{POPL '89}}. \bibinfo{publisher}{ACM},
  \bibinfo{pages}{60--76}.
\newblock


\bibitem[\protect\citeauthoryear{Yorgey, Weirich, Cretin, Peyton~Jones,
  Vytiniotis, and Magalh\~{a}es}{Yorgey et~al\mbox{.}}{2012}]%
        {hspromoted}
\bibfield{author}{\bibinfo{person}{Brent~A. Yorgey}, \bibinfo{person}{Stephanie
  Weirich}, \bibinfo{person}{Julien Cretin}, \bibinfo{person}{Simon
  Peyton~Jones}, \bibinfo{person}{Dimitrios Vytiniotis}, {and}
  \bibinfo{person}{Jos{\'e}~Pedro Magalh\~{a}es}.}
  \bibinfo{year}{2012}\natexlab{}.
\newblock \showarticletitle{Giving {Haskell} a Promotion}. In
  \bibinfo{booktitle}{\emph{TLDI '12}}. \bibinfo{publisher}{ACM},
  \bibinfo{pages}{53--66}.
\newblock


\end{thebibliography}

\end{document}